
%
\catcode`@=11 
%
%
%

\font\fourteenrm=cmr10 scaled\magstep2
\font\twelverm=cmr10 scaled\magstep1
\font\ninerm=cmr9            \font\sixrm=cmr6

\font\fourteenbf=cmbx10 scaled\magstep2
\font\twelvebf=cmbx10 scaled\magstep1
\font\ninebf=cmbx9            \font\sixbf=cmbx6
\font\seventeeni=cmmi10 scaled\magstep3     \skewchar\seventeeni='177
\font\fourteeni=cmmi10 scaled\magstep2      \skewchar\fourteeni='177
\font\twelvei=cmmi10 scaled\magstep1        \skewchar\twelvei='177
\font\ninei=cmmi9                           \skewchar\ninei='177
\font\sixi=cmmi6                            \skewchar\sixi='177
\font\seventeensy=cmsy10 scaled\magstep3    \skewchar\seventeensy='60
\font\fourteensy=cmsy10 scaled\magstep2     \skewchar\fourteensy='60
\font\twelvesy=cmsy10 scaled\magstep1       \skewchar\twelvesy='60
\font\ninesy=cmsy9                          \skewchar\ninesy='60
\font\sixsy=cmsy6                           \skewchar\sixsy='60

\font\fourteenex=cmex10 scaled\magstep2
\font\twelveex=cmex10 scaled\magstep1

\font\fourteensl=cmsl10 scaled\magstep2
\font\twelvesl=cmsl10 scaled\magstep1
\font\ninesl=cmsl9

\font\fourteenit=cmti10 scaled\magstep2
\font\twelveit=cmti10 scaled\magstep1
\font\twelvett=cmtt10 scaled\magstep1
\font\twelvecp=cmcsc10 scaled\magstep1
\font\tencp=cmcsc10
\newfam\cpfam
%
%
\newcount\f@ntkey            \f@ntkey=0
\def\samef@nt{\relax \ifcase\f@ntkey \rm \or\oldstyle \or\or
         \or\it \or\sl \or\bf \or\tt \or\caps \fi }
\def\fourteenpoint{\relax
    \textfont0=\fourteenrm          \scriptfont0=\tenrm
    \scriptscriptfont0=\sevenrm
     \def\rm{\fam0 \fourteenrm \f@ntkey=0 }\relax
    \textfont1=\fourteeni           \scriptfont1=\teni
    \scriptscriptfont1=\seveni
     \def\oldstyle{\fam1 \fourteeni\f@ntkey=1 }\relax
    \textfont2=\fourteensy          \scriptfont2=\tensy
    \scriptscriptfont2=\sevensy
    \textfont3=\fourteenex     \scriptfont3=\fourteenex
    \scriptscriptfont3=\fourteenex
    \def\it{\fam\itfam \fourteenit\f@ntkey=4 }\textfont\itfam=\fourteenit
    \def\sl{\fam\slfam \fourteensl\f@ntkey=5 }\textfont\slfam=\fourteensl
    \scriptfont\slfam=\tensl
    \def\bf{\fam\bffam \fourteenbf\f@ntkey=6 }\textfont\bffam=\fourteenbf
    \scriptfont\bffam=\tenbf     \scriptscriptfont\bffam=\sevenbf
    \def\tt{\fam\ttfam \twelvett \f@ntkey=7 }\textfont\ttfam=\twelvett
    \h@big=11.9\p@{} \h@Big=16.1\p@{} \h@bigg=20.3\p@{} \h@Bigg=24.5\p@{}
    \def\caps{\fam\cpfam \twelvecp \f@ntkey=8 }\textfont\cpfam=\twelvecp
    \setbox\strutbox=\hbox{\vrule height 12pt depth 5pt width\z@}
    \samef@nt}
\def\twelvepoint{\relax
    \textfont0=\twelverm          \scriptfont0=\ninerm
    \scriptscriptfont0=\sixrm
     \def\rm{\fam0 \twelverm \f@ntkey=0 }\relax
    \textfont1=\twelvei           \scriptfont1=\ninei
    \scriptscriptfont1=\sixi
     \def\oldstyle{\fam1 \twelvei\f@ntkey=1 }\relax
    \textfont2=\twelvesy          \scriptfont2=\ninesy
    \scriptscriptfont2=\sixsy
    \textfont3=\twelveex          \scriptfont3=\twelveex
    \scriptscriptfont3=\twelveex
    \def\it{\fam\itfam \twelveit \f@ntkey=4 }\textfont\itfam=\twelveit
    \def\sl{\fam\slfam \twelvesl \f@ntkey=5 }\textfont\slfam=\twelvesl
    \scriptfont\slfam=\ninesl
    \def\bf{\fam\bffam \twelvebf \f@ntkey=6 }\textfont\bffam=\twelvebf
    \scriptfont\bffam=\ninebf     \scriptscriptfont\bffam=\sixbf
    \def\tt{\fam\ttfam \twelvett \f@ntkey=7 }\textfont\ttfam=\twelvett
    \h@big=10.2\p@{}
    \h@Big=13.8\p@{}
    \h@bigg=17.4\p@{}
    \h@Bigg=21.0\p@{}
    \def\caps{\fam\cpfam \twelvecp \f@ntkey=8 }\textfont\cpfam=\twelvecp
    \setbox\strutbox=\hbox{\vrule height 10pt depth 4pt width\z@}
    \samef@nt}
\def\tenpoint{\relax
    \textfont0=\tenrm          \scriptfont0=\sevenrm
    \scriptscriptfont0=\fiverm
    \def\rm{\fam0 \tenrm \f@ntkey=0 }\relax
    \textfont1=\teni           \scriptfont1=\seveni
    \scriptscriptfont1=\fivei
    \def\oldstyle{\fam1 \teni \f@ntkey=1 }\relax
    \textfont2=\tensy          \scriptfont2=\sevensy
    \scriptscriptfont2=\fivesy
    \textfont3=\tenex          \scriptfont3=\tenex
    \scriptscriptfont3=\tenex
    \def\it{\fam\itfam \tenit \f@ntkey=4 }\textfont\itfam=\tenit
    \def\sl{\fam\slfam \tensl \f@ntkey=5 }\textfont\slfam=\tensl
    \def\bf{\fam\bffam \tenbf \f@ntkey=6 }\textfont\bffam=\tenbf
    \scriptfont\bffam=\sevenbf     \scriptscriptfont\bffam=\fivebf
    \def\tt{\fam\ttfam \tentt \f@ntkey=7 }\textfont\ttfam=\tentt
    \def\caps{\fam\cpfam \tencp \f@ntkey=8 }\textfont\cpfam=\tencp
    \setbox\strutbox=\hbox{\vrule height 8.5pt depth 3.5pt width\z@}
    \samef@nt}
%
%
%
%
\newdimen\h@big  \h@big=8.5\p@
\newdimen\h@Big  \h@Big=11.5\p@
\newdimen\h@bigg  \h@bigg=14.5\p@
\newdimen\h@Bigg  \h@Bigg=17.5\p@
\def\big#1{{\hbox{$\left#1\vbox to\h@big{}\right.\n@space$}}}
\def\Big#1{{\hbox{$\left#1\vbox to\h@Big{}\right.\n@space$}}}
\def\bigg#1{{\hbox{$\left#1\vbox to\h@bigg{}\right.\n@space$}}}
\def\Bigg#1{{\hbox{$\left#1\vbox to\h@Bigg{}\right.\n@space$}}}
%
%
%
\normalbaselineskip = 20pt plus 0.2pt minus 0.1pt
\normallineskip = 1.5pt plus 0.1pt minus 0.1pt
\normallineskiplimit = 1.5pt
\newskip\normaldisplayskip
\normaldisplayskip = 20pt plus 5pt minus 10pt
\newskip\normaldispshortskip
\normaldispshortskip = 6pt plus 5pt
\newskip\normalparskip
\normalparskip = 6pt plus 2pt minus 1pt
\newskip\skipregister
\skipregister = 5pt plus 2pt minus 1.5pt
\newif\ifsingl@    \newif\ifdoubl@
\newif\iftwelv@    \twelv@true
\def\singlespace{\singl@true\doubl@false\spaces@t}
\def\doublespace{\singl@false\doubl@true\spaces@t}
\def\normalspace{\singl@false\doubl@false\spaces@t}
\def\Tenpoint{\tenpoint\twelv@false\spaces@t}
\def\Twelvepoint{\twelvepoint\twelv@true\spaces@t}
\def\spaces@t{\relax%
 \iftwelv@ \ifsingl@\subspaces@t3:4;\else\subspaces@t29:31;\fi%
 \else \ifsingl@\subspaces@t3:5;\else\subspaces@t4:5;\fi \fi%
 \ifdoubl@ \multiply\baselineskip by 5%
 \divide\baselineskip by 4 \fi \unskip}
\def\subspaces@t#1:#2;{
      \baselineskip = \normalbaselineskip
      \multiply\baselineskip by #1 \divide\baselineskip by #2
      \lineskip = \normallineskip
      \multiply\lineskip by #1 \divide\lineskip by #2
      \lineskiplimit = \normallineskiplimit
      \multiply\lineskiplimit by #1 \divide\lineskiplimit by #2
      \parskip = \normalparskip
      \multiply\parskip by #1 \divide\parskip by #2
      \abovedisplayskip = \normaldisplayskip
      \multiply\abovedisplayskip by #1 \divide\abovedisplayskip by #2
      \belowdisplayskip = \abovedisplayskip
      \abovedisplayshortskip = \normaldispshortskip
      \multiply\abovedisplayshortskip by #1
        \divide\abovedisplayshortskip by #2
      \belowdisplayshortskip = \abovedisplayshortskip
      \advance\belowdisplayshortskip by \belowdisplayskip
      \divide\belowdisplayshortskip by 2
      \smallskipamount = \skipregister
      \multiply\smallskipamount by #1 \divide\smallskipamount by #2
      \medskipamount = \smallskipamount \multiply\medskipamount by 2
      \bigskipamount = \smallskipamount \multiply\bigskipamount by 4 }
\def\normalbaselines{ \baselineskip=\normalbaselineskip
   \lineskip=\normallineskip \lineskiplimit=\normallineskip
   \iftwelv@\else \multiply\baselineskip by 4 \divide\baselineskip by 5
     \multiply\lineskiplimit by 4 \divide\lineskiplimit by 5
     \multiply\lineskip by 4 \divide\lineskip by 5 \fi }
\Twelvepoint  
\interlinepenalty=50
\interfootnotelinepenalty=5000
\predisplaypenalty=9000
\postdisplaypenalty=500
\hfuzz=1pt
\vfuzz=0.2pt
%
%
%
\def\pagecontents{
   \ifvoid\topins\else\unvbox\topins\vskip\skip\topins\fi
   \dimen@ = \dp255 \unvbox255
   \ifvoid\footins\else\vskip\skip\footins\footrule\unvbox\footins\fi
   \ifr@ggedbottom \kern-\dimen@ \vfil \fi }
\def\makeheadline{\vbox to 0pt{ \skip@=\topskip
      \advance\skip@ by -12pt \advance\skip@ by -2\normalbaselineskip
      \vskip\skip@ \line{\vbox to 12pt{}\the\headline} \vss
      }\nointerlineskip}
\def\makefootline{\baselineskip = 1.5\normalbaselineskip
                 \line{\the\footline}}
\newif\iffrontpage
\newif\ifletterstyle
\newif\ifp@genum
\def\nopagenumbers{\p@genumfalse}
\def\pagenumbers{\p@genumtrue}
\pagenumbers
\newtoks\paperheadline
\newtoks\letterheadline
\newtoks\letterfrontheadline
\newtoks\lettermainheadline
\newtoks\paperfootline
\newtoks\letterfootline
\newtoks\date
\footline={\ifletterstyle\the\letterfootline\else\the\paperfootline\fi}
\paperfootline={\hss\iffrontpage\else\ifp@genum\tenrm\folio\hss\fi\fi}
\letterfootline={\hfil}
\headline={\ifletterstyle\the\letterheadline\else\the\paperheadline\fi}
\paperheadline={\hfil}
\letterheadline{\iffrontpage\the\letterfrontheadline
     \else\the\lettermainheadline\fi}
\lettermainheadline={\rm\ifp@genum page \ \folio\fi\hfil\the\date}
\def\monthname{\relax\ifcase\month 0/\or January\or February\or
   March\or April\or May\or June\or July\or August\or September\or
   October\or November\or December\else\number\month/\fi}
\date={\monthname\ \number\day, \number\year}
\countdef\pagenumber=1  \pagenumber=1
\def\advancepageno{\global\advance\pageno by 1
   \ifnum\pagenumber<0 \global\advance\pagenumber by -1
    \else\global\advance\pagenumber by 1 \fi \global\frontpagefalse }
\def\folio{\ifnum\pagenumber<0 \romannumeral-\pagenumber
           \else \number\pagenumber \fi }
\def\footrule{\dimen@=\prevdepth\nointerlineskip
   \vbox to 0pt{\vskip -0.25\baselineskip \hrule width 0.35\hsize \vss}
   \prevdepth=\dimen@ }
\newtoks\foottokens
\foottokens={\Tenpoint\singlespace}
\newdimen\footindent
\footindent=24pt
\def\vfootnote#1{\insert\footins\bgroup  \the\foottokens
   \interlinepenalty=\interfootnotelinepenalty \floatingpenalty=20000
   \splittopskip=\ht\strutbox \boxmaxdepth=\dp\strutbox
   \leftskip=\footindent \rightskip=\z@skip
   \parindent=0.5\footindent \parfillskip=0pt plus 1fil
   \spaceskip=\z@skip \xspaceskip=\z@skip
   \Textindent{$ #1 $}\footstrut\futurelet\next\fo@t}
\def\Textindent#1{\noindent\llap{#1\enspace}\ignorespaces}
\def\footnote#1{\attach{#1}\vfootnote{#1}}

\let\footsymbol=\star
\newcount\lastf@@t           \lastf@@t=-1
\newcount\footsymbolcount    \footsymbolcount=0
\newif\ifPhysRev
\def\footsymbolgen{\relax \ifPhysRev \iffrontpage \NPsymbolgen\else
      \PRsymbolgen\fi \else \NPsymbolgen\fi
   \global\lastf@@t=\pageno \footsymbol }
\def\NPsymbolgen{\ifnum\footsymbolcount<0 \global\footsymbolcount=0\fi
   {\iffrontpage \else \advance\lastf@@t by 1 \fi
    \ifnum\lastf@@t<\pageno \global\footsymbolcount=0
     \else \global\advance\footsymbolcount by 1 \fi }
   \ifcase\footsymbolcount \fd@f\star\or \fd@f\dagger\or \fd@f\ast\or
    \fd@f\ddagger\or \fd@f\natural\or \fd@f\diamond\or \fd@f\bullet\or
    \fd@f\nabla\else \fd@f\dagger\global\footsymbolcount=0 \fi }
\def\fd@f#1{\xdef\footsymbol{#1}}
\def\PRsymbolgen{\ifnum\footsymbolcount>0 \global\footsymbolcount=0\fi
      \global\advance\footsymbolcount by -1
      \xdef\footsymbol{\sharp\number-\footsymbolcount} }
\def\space@ver#1{\let\@sf=\empty \ifmmode #1\else \ifhmode
   \edef\@sf{\spacefactor=\the\spacefactor}\unskip${}#1$\relax\fi\fi}
\def\attach#1{\space@ver{\strut^{\mkern 2mu #1} }\@sf\ }
\def\atttach#1{\space@ver{\strut{\mkern 2mu #1} }\@sf\ }
%
%
%
\newcount\chapternumber      \chapternumber=0
\newcount\sectionnumber      \sectionnumber=0
\newcount\equanumber         \equanumber=0
\let\chapterlabel=0
\newtoks\chapterstyle        \chapterstyle={\Number}
\newskip\chapterskip         \chapterskip=\bigskipamount
\newskip\sectionskip         \sectionskip=\medskipamount
\newskip\headskip            \headskip=8pt plus 3pt minus 3pt
\newdimen\chapterminspace    \chapterminspace=15pc
\newdimen\sectionminspace    \sectionminspace=10pc
\newdimen\referenceminspace  \referenceminspace=25pc
\def\chapterreset{\global\advance\chapternumber by 1
   \ifnum\the\equanumber<0 \else\global\equanumber=0\fi
   \sectionnumber=0 \makel@bel}
\def\makel@bel{\xdef\chapterlabel{%
\the\chapterstyle{\the\chapternumber}.}}
\def\sectionlabel{\number\sectionnumber \quad }
\def\alphabetic#1{\count255='140 \advance\count255 by #1\char\count255}
\def\Alphabetic#1{\count255='100 \advance\count255 by #1\char\count255}
\def\Roman#1{\uppercase\expandafter{\romannumeral #1}}
\def\roman#1{\romannumeral #1}
\def\Number#1{\number #1}
\def\unnumberedchapters{\let\makel@bel=\relax \let\chapterlabel=\relax
\let\sectionlabel=\relax \equanumber=-1 }
\def\titlestyle#1{\par\begingroup \interlinepenalty=9999
     \leftskip=0.02\hsize plus 0.23\hsize minus 0.02\hsize
     \rightskip=\leftskip \parfillskip=0pt
     \hyphenpenalty=9000 \exhyphenpenalty=9000
     \tolerance=9999 \pretolerance=9000
     \spaceskip=0.333em \xspaceskip=0.5em
     \iftwelv@\fourteenpoint\else\twelvepoint\fi
   \noindent #1\par\endgroup }
\def\spacecheck#1{\dimen@=\pagegoal\advance\dimen@ by -\pagetotal
   \ifdim\dimen@<#1 \ifdim\dimen@>0pt \vfil\break \fi\fi}
\def\chapter#1{\par \penalty-300 \vskip\chapterskip
   \spacecheck\chapterminspace
   \chapterreset \titlestyle{\chapterlabel \ #1}
   \nobreak\vskip\headskip \penalty 30000
   \wlog{\string\chapter\ \chapterlabel} }

\def\section#1{\par \ifnum\the\lastpenalty=30000\else
   \penalty-200\vskip\sectionskip \spacecheck\sectionminspace\fi
   \wlog{\string\section\ \chapterlabel \the\sectionnumber}
   \global\advance\sectionnumber by 1  \noindent
   {\caps\enspace\chapterlabel \sectionlabel #1}\par
   \nobreak\vskip\headskip \penalty 30000 }
\def\subsection#1{\par
   \ifnum\the\lastpenalty=30000\else \penalty-100\smallskip \fi
   \noindent\undertext{#1}\enspace \vadjust{\penalty5000}}

\def\undertext#1{\vtop{\hbox{#1}\kern 1pt \hrule}}
\def\APPENDIX#1#2{\par\penalty-300\vskip\chapterskip
   \spacecheck\chapterminspace \chapterreset \xdef\chapterlabel{#1}
   \titlestyle{APPENDIX #2} \nobreak\vskip\headskip \penalty 30000
   \wlog{\string\Appendix\ \chapterlabel} }
\def\Appendix#1{\APPENDIX{#1}{#1}}
\def\appendix{\APPENDIX{A}{}}
%
%
%
\def\eqname#1{\relax \ifnum\the\equanumber<0
     \xdef#1{{\rm(\number-\equanumber)}}\global\advance\equanumber by -1
    \else \global\advance\equanumber by 1
      \xdef#1{{\rm(\chapterlabel \number\equanumber)}} \fi}

\def\eqn#1{\eqno\eqname{#1}#1}

\def\eqinsert#1{\noalign{\dimen@=\prevdepth \nointerlineskip
   \setbox0=\hbox to\displaywidth{\hfil #1}
   \vbox to 0pt{\vss\hbox{$\!\box0\!$}\kern-0.5\baselineskip}
   \prevdepth=\dimen@}}
\def\sequentialequations{\globaleqnumbers}
%
%
\def\GENITEM#1;#2{\par \hangafter=0 \hangindent=#1
    \Textindent{$ #2 $}\ignorespaces}
\outer\def\newitem#1=#2;{\gdef#1{\GENITEM #2;}}
\newdimen\itemsize                \itemsize=30pt
\newitem\item=1\itemsize;
\newitem\sitem=1.75\itemsize;     
\newitem\ssitem=2.5\itemsize;     
\outer\def\newlist#1=#2&#3&#4;{\toks0={#2}\toks1={#3}%
   \count255=\escapechar \escapechar=-1
   \alloc@0\list\countdef\insc@unt\listcount     \listcount=0
   \edef#1{\par
      \countdef\listcount=\the\allocationnumber
      \advance\listcount by 1
      \hangafter=0 \hangindent=#4
      \Textindent{\the\toks0{\listcount}\the\toks1}}
   \expandafter\expandafter\expandafter
    \edef\c@t#1{begin}{\par
      \countdef\listcount=\the\allocationnumber \listcount=1
      \hangafter=0 \hangindent=#4
      \Textindent{\the\toks0{\listcount}\the\toks1}}
   \expandafter\expandafter\expandafter
    \edef\c@t#1{con}{\par \hangafter=0 \hangindent=#4 \noindent}
   \escapechar=\count255}
\def\c@t#1#2{\csname\string#1#2\endcsname}
\newlist\point=\Number&.&1.0\itemsize;
\newlist\subpoint=(\alphabetic&)&1.75\itemsize;
\newlist\subsubpoint=(\roman&)&2.5\itemsize;
\newlist\cpoint=\Roman&.&1.0\itemsize;
%

%
%
%
\newcount\referencecount     \referencecount=0
\newif\ifreferenceopen       \newwrite\referencewrite
\newtoks\rw@toks
\def\NPrefmark#1{\atttach{\rm [ #1 ] }}
\let\PRrefmark=\attach
\def\CErefmark#1{\attach{\scriptstyle  #1 ) }}
\def\refmark#1{\relax\ifPhysRev\PRrefmark{#1}\else\NPrefmark{#1}\fi}
\def\crefmark#1{\relax\CErefmark{#1}}
\def\refend{\refmark{\number\referencecount}}
\newcount\lastrefsbegincount \lastrefsbegincount=0
\def\refsend{\refmark{\count255=\referencecount
   \advance\count255 by-\lastrefsbegincount
   \ifcase\count255 \number\referencecount
   \or \number\lastrefsbegincount,\number\referencecount
   \else \number\lastrefsbegincount-\number\referencecount \fi}}
\def\crefsend{\crefmark{\count255=\referencecount
   \advance\count255 by-\lastrefsbegincount
   \ifcase\count255 \number\referencecount
   \or \number\lastrefsbegincount,\number\referencecount
   \else \number\lastrefsbegincount-\number\referencecount \fi}}
\def\refch@ck{\chardef\rw@write=\referencewrite
   \ifreferenceopen \else \referenceopentrue
   \immediate\openout\referencewrite=referenc.texauxil \fi}
%
{\catcode`\^^M=\active 
  \gdef\obeyendofline{\catcode`\^^M\active \let^^M\ }}%
%
{\catcode`\^^M=\active 
  \gdef\ignoreendofline{\catcode`\^^M=5}}
{\obeyendofline\gdef\rw@start#1{\def\t@st{#1} \ifx\t@st\blankend%
\endgroup \@sf \relax \else \ifx\t@st\bl@nkend \endgroup \@sf \relax%
\else \rw@begin#1
\backtotext
\fi \fi } }
{\obeyendofline\gdef\rw@begin#1
{\def\n@xt{#1}\rw@toks={#1}\relax%
\rw@next}}
\def\blankend{}
{\obeylines\gdef\bl@nkend{
}}
\newif\iffirstrefline  \firstreflinetrue
\def\rwr@teswitch{\ifx\n@xt\blankend \let\n@xt=\rw@begin %
 \else\iffirstrefline \global\firstreflinefalse%
\immediate\write\rw@write{\noexpand\obeyendofline \the\rw@toks}%
\let\n@xt=\rw@begin%
      \else\ifx\n@xt\rw@@d \def\n@xt{\immediate\write\rw@write{%
        \noexpand\ignoreendofline}\endgroup \@sf}%
             \else \immediate\write\rw@write{\the\rw@toks}%
             \let\n@xt=\rw@begin\fi\fi \fi}
\def\rw@next{\rwr@teswitch\n@xt}
\def\rw@@d{\backtotext} \let\rw@end=\relax
\let\backtotext=\relax

\newdimen\refindent     \refindent=30pt
\def\refitem#1{\par \hangafter=0 \hangindent=\refindent \Textindent{#1}}
\def\REFNUM#1{\space@ver{}\refch@ck \firstreflinetrue%
 \global\advance\referencecount by 1 \xdef#1{\the\referencecount}}
\def\refnum#1{\space@ver{}\refch@ck \firstreflinetrue%
 \global\advance\referencecount by 1 \xdef#1{\the\referencecount}\refend}

\def\REF#1{\REFNUM#1%
 \immediate\write\referencewrite{%
 \noexpand\refitem{#1.}}%
\begingroup\obeyendofline\rw@start}
\def\ref{\refnum\?%
 \immediate\write\referencewrite{\noexpand\refitem{\?.}}%
\begingroup\obeyendofline\rw@start}
\def\Ref#1{\refnum#1%
 \immediate\write\referencewrite{\noexpand\refitem{#1.}}%
\begingroup\obeyendofline\rw@start}
\def\REFS#1{\REFNUM#1\global\lastrefsbegincount=\referencecount
\immediate\write\referencewrite{\noexpand\refitem{#1.}}%
\begingroup\obeyendofline\rw@start}
\newcount\figurecount     \figurecount=0
\newif\iffigureopen       \newwrite\figurewrite
\def\figch@ck{\chardef\rw@write=\figurewrite \iffigureopen\else
   \immediate\openout\figurewrite=figures.texauxil
   \figureopentrue\fi}
\def\FIGNUM#1{\space@ver{}\figch@ck \firstreflinetrue%
 \global\advance\figurecount by 1 \xdef#1{\the\figurecount}}
\def\FIG#1{\FIGNUM#1
   \immediate\write\figurewrite{\noexpand\refitem{#1.}}%
   \begingroup\obeyendofline\rw@start}
\def\fig{\FIGNUM\? fig.~\?%
\immediate\write\figurewrite{\noexpand\refitem{\?.}}%
\begingroup\obeyendofline\rw@start}
\def\figure{\FIGNUM\? figure~\?
   \immediate\write\figurewrite{\noexpand\refitem{\?.}}%
   \begingroup\obeyendofline\rw@start}
\def\Fig{\FIGNUM\? Fig.~\?%
\immediate\write\figurewrite{\noexpand\refitem{\?.}}%
\begingroup\obeyendofline\rw@start}
\def\Figure{\FIGNUM\? Figure~\?%
\immediate\write\figurewrite{\noexpand\refitem{\?.}}%
\begingroup\obeyendofline\rw@start}
\newcount\tablecount     \tablecount=0
\newif\iftableopen       \newwrite\tablewrite
\def\tabch@ck{\chardef\rw@write=\tablewrite \iftableopen\else
   \immediate\openout\tablewrite=tables.texauxil
   \tableopentrue\fi}
\def\TABNUM#1{\space@ver{}\tabch@ck \firstreflinetrue%
 \global\advance\tablecount by 1 \xdef#1{\the\tablecount}}
\def\TABLE#1{\TABNUM#1
   \immediate\write\tablewrite{\noexpand\refitem{#1.}}%
   \begingroup\obeyendofline\rw@start}
\def\Table{\TABNUM\? Table~\?%
\immediate\write\tablewrite{\noexpand\refitem{\?.}}%
\begingroup\obeyendofline\rw@start}
%

%
%
%
\def\masterreset{\global\pagenumber=1 \global\chapternumber=0
   \ifnum\the\equanumber<0\else \global\equanumber=0\fi
   \global\sectionnumber=0
   \global\referencecount=0 \global\figurecount=0 \global\tablecount=0 }
\def\FRONTPAGE{\ifvoid255\else\vfill\penalty-2000\fi
      \masterreset\global\frontpagetrue
      \global\lastf@@t=0 \global\footsymbolcount=0}

\def\paperstyle{\letterstylefalse\normalspace\papersize}
\def\letterstyle{\letterstyletrue\singlespace\lettersize}
\def\papersize{\hsize=6.5truein\vsize=9.1truein\hoffset=-.3truein
               \voffset=-.4truein\skip\footins=\bigskipamount}
\def\lettersize{\hsize=6.5truein\vsize=9.1truein\hoffset=-.3truein
    \voffset=.1truein\skip\footins=\smallskipamount \multiply
    \skip\footins by 3 }
\paperstyle   
%
%
\def\MEMO{\letterstyle\FRONTPAGE \letterfrontheadline={\hfil}
    \line{\quad\fourteenrm CERN MEMORANDUM\hfil\twelverm\the\date\quad}
    \medskip \memod@f}

\def\memit@m#1{\smallskip \hangafter=0 \hangindent=1in
      \Textindent{\caps #1}}
\def\memod@f{\xdef\mto{\memit@m{To:}}\xdef\from{\memit@m{From:}}%
     \xdef\topic{\memit@m{Topic:}}\xdef\subject{\memit@m{Subject:}}%
     \xdef\rule{\bigskip\hrule height 1pt\bigskip}}
\memod@f
\newskip\lettertopfil
\lettertopfil = 0pt plus 1.5in minus 0pt
\newskip\letterbottomfil
\letterbottomfil = 0pt plus 2.3in minus 0pt
\newskip\spskip \setbox0\hbox{\ } \spskip=-1\wd0
\def\addressee#1{\medskip\rightline{\the\date\hskip 30pt} \bigskip
   \vskip\lettertopfil
   \ialign to\hsize{\strut ##\hfil\tabskip 0pt plus \hsize \cr #1\crcr}
   \medskip\noindent\hskip\spskip}
\newskip\signatureskip       \signatureskip=40pt
\def\signed#1{\par \penalty 9000 \bigskip \dt@pfalse
  \everycr={\noalign{\ifdt@p\vskip\signatureskip\global\dt@pfalse\fi}}
  \setbox0=\vbox{\singlespace \halign{\tabskip 0pt \strut ##\hfil\cr
   \noalign{\global\dt@ptrue}#1\crcr}}
  \line{\hskip 0.5\hsize minus 0.5\hsize \box0\hfil} \medskip }

\def\endletter{\ifnum\pagenumber=1 \vskip\letterbottomfil\supereject
\else \vfil\supereject \fi}
\newbox\letterb@x
\def\lettertext{\par\unvcopy\letterb@x\par}
\def\multiletter{\setbox\letterb@x=\vbox\bgroup
      \everypar{\vrule height 1\baselineskip depth 0pt width 0pt }
      \singlespace \topskip=\baselineskip }
\def\letterend{\par\egroup}
%
%
%
\newskip\frontpageskip
\newtoks\pubtype
\newtoks\Pubnum
\newtoks\pubnum
\newtoks\pubnu
\newtoks\pubn
\newif\ifp@bblock  \p@bblocktrue
\def\PH@SR@V{\doubl@true \baselineskip=24.1pt plus 0.2pt minus 0.1pt
             \parskip= 3pt plus 2pt minus 1pt }
\def\PHYSREV{\paperstyle\PhysRevtrue\PH@SR@V}
\def\titlepage{\FRONTPAGE\paperstyle\ifPhysRev\PH@SR@V\fi
   \ifp@bblock\p@bblock\fi}
\def\nopubblock{\p@bblockfalse}
\def\endpage{\vfil\break}
\frontpageskip=1\medskipamount plus .5fil
\pubtype={\tensl Preliminary Version}
\Pubnum={$\rm CERN-TH.\the\pubnum $}
\pubnum={0000}
\def\p@bblock{\begingroup \tabskip=\hsize minus \hsize
   \baselineskip=1.5\ht\strutbox \topspace-2\baselineskip
   \halign to\hsize{\strut ##\hfil\tabskip=0pt\crcr
   \the \pubn\cr
   \the \Pubnum\cr
   \the \pubnu\cr
   \the \date\cr}\endgroup}
\def\title#1{\vskip\frontpageskip \titlestyle{#1} \vskip\headskip }
\def\author#1{\vskip\frontpageskip\titlestyle{\twelvecp #1}\nobreak}

\def\address#1{\par\kern 5pt\titlestyle{\twelvepoint\it #1}}
\def\andaddress{\par\kern 5pt \centerline{\sl and} \address}

\def\abstract{\vskip\frontpageskip\centerline{\fourteenrm ABSTRACT}
              \vskip\headskip }

%
%
%

\def\\{\relax\ifmmode\backslash\else$\backslash$\fi}
\def\globaleqnumbers{\relax\ifnum\the\equanumber<0%
\else\global\equanumber=-1\fi}
\def\nextline{\unskip\nobreak\hskip\parfillskip\break}

\def\journal#1&#2(#3){\unskip, \sl #1~\bf #2 \rm (19#3) }

\def\topspace{\hrule height 0pt depth 0pt \vskip}

\let\int=\intop         
\def\prop{\mathrel{{\mathchoice{\pr@p\scriptstyle}{\pr@p\scriptstyle}{
                \pr@p\scriptscriptstyle}{\pr@p\scriptscriptstyle} }}}
\def\pr@p#1{\setbox0=\hbox{$\cal #1 \char'103$}
   \hbox{$\cal #1 \char'117$\kern-.4\wd0\box0}}
\def\lsim{\mathrel{\mathpalette\@versim<}}
\def\gsim{\mathrel{\mathpalette\@versim>}}
\def\@versim#1#2{\lower0.2ex\vbox{\baselineskip\z@skip\lineskip\z@skip
  \lineskiplimit\z@\ialign{$\m@th#1\hfil##\hfil$\crcr#2\crcr\sim\crcr}}}
\def\leftrightarrowfill{$\m@th \mathord- \mkern-6mu
        \cleaders\hbox{$\mkern-2mu \mathord- \mkern-2mu$}\hfil
        \mkern-6mu \mathord\leftrightarrow$}
\def\lrover#1{\vbox{\ialign{##\crcr
        \leftrightarrowfill\crcr\noalign{\kern-1pt\nointerlineskip}
        $\hfil\displaystyle{#1}\hfil$\crcr}}}
%
%
%
\let\sec@nt=\sec
\def\sec{\relax\ifmmode\let\n@xt=\sec@nt\else\let\n@xt\section\fi\n@xt}
\def\obsolete#1{\message{Macro \string #1 is obsolete.}}
\def\firstsec#1{\obsolete\firstsec \section{#1}}
\def\firstsubsec#1{\obsolete\firstsubsec \subsection{#1}}
\def\thispage#1{\obsolete\thispage \global\pagenumber=#1\frontpagefalse}
\def\thischapter#1{\obsolete\thischapter \global\chapternumber=#1}
\def\nextequation#1{\obsolete\nextequation \global\equanumber=#1
   \ifnum\the\equanumber>0 \global\advance\equanumber by 1 \fi}
\def\BOXITEM{\afterassigment\B@XITEM\setbox0=}
\def\B@XITEM{\par\hangindent\wd0 \noindent\box0 }
%

%
%

%
%

%
%

%

%

%

%

%

%
%
%
\def\boxit#1{\vbox{\hrule\hbox{\vrule\kern3pt\vbox{\kern3pt#1\kern3pt}
\kern3pt\vrule}\hrule}}
%
%
%
\catcode`@=12 
\message{ by V.K./U.B.}
\everyjob{\input imyphyx }
%
%
\newbox\hdbox%
\newcount\hdrows%
\newcount\multispancount%
\newcount\ncase%
\newcount\ncols
\newcount\nrows%
\newcount\nspan%
\newcount\ntemp%
\newdimen\hdsize%
\newdimen\newhdsize%
\newdimen\parasize%
\newdimen\spreadwidth%
\newdimen\thicksize%
\newdimen\thinsize%
\newdimen\tablewidth%
\newif\ifcentertables%
\newif\ifendsize%
\newif\iffirstrow%
\newif\iftableinfo%
\newtoks\dbt%
\newtoks\hdtks%
\newtoks\savetks%
\newtoks\tableLETtokens%
\newtoks\tabletokens%
\newtoks\widthspec%
%
%
%
%
\tableinfotrue%
\catcode`\@=11
%
%
\def\tstrut{\vrule height3.1ex depth1.2ex width0pt}%
\def\and{\char`\&}
\def\tablerule{\noalign{\hrule height\thinsize depth0pt}}%
\thicksize=1.5pt
\thinsize=0.6pt
\def\thickrule{\noalign{\hrule height\thicksize depth0pt}}%
\def\ctr#1{\hfil\ #1\hfil}%
%
%
%
%
\tablewidth=-\maxdimen%
\spreadwidth=-\maxdimen%
\def\tabskipglue{0pt plus 1fil minus 1fil}%
%
%
\centertablestrue%
%
%
%
%
\parasize=4in%
\gdef\ARGS{########}
\gdef\headerARGS{####}
\def\@mpersand{&}
{\catcode`\|=13
\gdef\letbarzero{\let|0}
\gdef\letbartab{\def|{&&}}%
\gdef\letvbbar{\let\vb|}%
}
{\catcode`\&=4
\def\ampskip{&\omit\hfil&}
\catcode`\&=13
\let&0
\xdef\letampskip{\def&{\ampskip}}%
\gdef\letnovbamp{\let\novb&\let\tab&}
}
\def\begintable{
   \begingroup%
   \catcode`\|=13\letbartab\letvbbar%
   \catcode`\&=13\letampskip\letnovbamp%
   \def\multispan##1{
      \omit \mscount##1%
      \multiply\mscount\tw@\advance\mscount\m@ne%
      \loop\ifnum\mscount>\@ne \sp@n\repeat%
   }
   \def\|{%
      &\omit\widevline&%
   }%
   \ruledtable
}
\long\def\ruledtable#1\endtable{%
%
%
%
   \offinterlineskip
   \tabskip 0pt
   \def\widevline{\vrule width\thicksize}
   \def\endrow{\@mpersand\omit\hfil\crnorm\@mpersand}%
   \def\crthick{\@mpersand\crnorm\thickrule\@mpersand}%
   \def\crthickneg##1{\@mpersand\crnorm\thickrule
          \noalign{{\skip0=##1\vskip-\skip0}}\@mpersand}%
   \def\crnorule{\@mpersand\crnorm\@mpersand}%
   \def\crnoruleneg##1{\@mpersand\crnorm
          \noalign{{\skip0=##1\vskip-\skip0}}\@mpersand}%
   \let\nr=\crnorule
   \def\endtable{\@mpersand\crnorm\thickrule}%
   \let\crnorm=\cr
%
%
   \edef\cr{\@mpersand\crnorm\tablerule\@mpersand}%
   \def\crneg##1{\@mpersand\crnorm\tablerule
          \noalign{{\skip0=##1\vskip-\skip0}}\@mpersand}%
   \let\ctneg=\crthickneg
   \let\nrneg=\crnoruleneg
   \the\tableLETtokens
%
%
   \tabletokens={&#1}
%
%
   \countROWS\tabletokens\into\nrows%
   \countCOLS\tabletokens\into\ncols%
%
%
   \advance\ncols by -1%
   \divide\ncols by 2%
   \advance\nrows by 1%
%
%
   \iftableinfo %
      \immediate\write16{[Nrows=\the\nrows, Ncols=\the\ncols]}%
   \fi%
%
%
   \ifcentertables
      \ifhmode \par\fi
      \hbox to \hsize{
      \hss
   \else %
      \hbox{%
   \fi
      \vbox{%
         \makePREAMBLE{\the\ncols}
         \edef\next{\preamble}
         \let\preamble=\next
         \makeTABLE{\preamble}{\tabletokens}
      }
      \ifcentertables \hss}\else }\fi
   \endgroup
   \tablewidth=-\maxdimen
   \spreadwidth=-\maxdimen
}
\def\makeTABLE#1#2{
   {
   \let\ifmath0
   \let\header0
   \let\multispan0
%
%
   \ncase=0%
   \ifdim\tablewidth>-\maxdimen \ncase=1\fi%
   \ifdim\spreadwidth>-\maxdimen \ncase=2\fi%
   \relax
%
   \ifcase\ncase %
      \widthspec={}%
   \or %
      \widthspec=\expandafter{\expandafter t\expandafter o%
                 \the\tablewidth}%
   \else %
      \widthspec=\expandafter{\expandafter s\expandafter p\expandafter r%
                 \expandafter e\expandafter a\expandafter d%
                 \the\spreadwidth}%
   \fi %
   \xdef\next{
      \halign\the\widthspec{%
      #1
      \noalign{\hrule height\thicksize depth0pt}
      \the#2\endtable
%
      }
   }
   }
   \next
}
\def\makePREAMBLE#1{
   \ncols=#1
   \begingroup
   \let\ARGS=0
   \edef\xtp{\widevline\ARGS\tabskip\tabskipglue%
   &\ctr{\ARGS}\tstrut}
   \advance\ncols by -1
   \loop
      \ifnum\ncols>0 %
      \advance\ncols by -1%
      \edef\xtp{\xtp&\vrule width\thinsize\ARGS&\ctr{\ARGS}}%
   \repeat
   \xdef\preamble{\xtp&\widevline\ARGS\tabskip0pt%
   \crnorm}
   \endgroup
}
\def\countROWS#1\into#2{
   \let\countREGISTER=#2%
   \countREGISTER=0%
   \expandafter\ROWcount\the#1\endcount%
}%
\def\ROWcount{%
   \afterassignment\subROWcount\let\next= %
}%
\def\subROWcount{%
   \ifx\next\endcount %
      \let\next=\relax%
   \else%
      \ncase=0%
      \ifx\next\cr %
         \global\advance\countREGISTER by 1%
         \ncase=0%
      \fi%
      \ifx\next\endrow %
         \global\advance\countREGISTER by 1%
         \ncase=0%
      \fi%
      \ifx\next\crthick %
         \global\advance\countREGISTER by 1%
         \ncase=0%
      \fi%
      \ifx\next\crnorule %
         \global\advance\countREGISTER by 1%
         \ncase=0%
      \fi%
      \ifx\next\crthickneg %
         \global\advance\countREGISTER by 1%
         \ncase=0%
      \fi%
      \ifx\next\crnoruleneg %
         \global\advance\countREGISTER by 1%
         \ncase=0%
      \fi%
      \ifx\next\crneg %
         \global\advance\countREGISTER by 1%
         \ncase=0%
      \fi%
      \ifx\next\header %
         \ncase=1%
      \fi%
      \relax%
      \ifcase\ncase %
         \let\next\ROWcount%
      \or %
         \let\next\argROWskip%
      \else %
      \fi%
   \fi%
   \next%
}
\def\counthdROWS#1\into#2{%
\dvr{10}%
   \let\countREGISTER=#2%
   \countREGISTER=0%
\dvr{11}%
\dvr{13}%
   \expandafter\hdROWcount\the#1\endcount%
\dvr{12}%
}%
\def\hdROWcount{%
   \afterassignment\subhdROWcount\let\next= %
}%
\def\subhdROWcount{%
   \ifx\next\endcount %
      \let\next=\relax%
   \else%
      \ncase=0%
      \ifx\next\cr %
         \global\advance\countREGISTER by 1%
         \ncase=0%
      \fi%
      \ifx\next\endrow %
         \global\advance\countREGISTER by 1%
         \ncase=0%
      \fi%
      \ifx\next\crthick %
         \global\advance\countREGISTER by 1%
         \ncase=0%
      \fi%
      \ifx\next\crnorule %
         \global\advance\countREGISTER by 1%
         \ncase=0%
      \fi%
      \ifx\next\header %
         \ncase=1%
      \fi%
\relax%
      \ifcase\ncase %
         \let\next\hdROWcount%
      \or%
         \let\next\arghdROWskip%
      \else %
      \fi%
   \fi%
   \next%
}%
{\catcode`\|=13\letbartab
\gdef\countCOLS#1\into#2{%
   \let\countREGISTER=#2%
   \global\countREGISTER=0%
   \global\multispancount=0%
   \global\firstrowtrue
   \expandafter\COLcount\the#1\endcount%
   \global\advance\countREGISTER by 3%
   \global\advance\countREGISTER by -\multispancount
}%
\gdef\COLcount{%
   \afterassignment\subCOLcount\let\next= %
}%
{\catcode`\&=13%
\gdef\subCOLcount{%
   \ifx\next\endcount %
      \let\next=\relax%
   \else%
      \ncase=0%
      \iffirstrow
         \ifx\next& %
            \global\advance\countREGISTER by 2%
            \ncase=0%
         \fi%
         \ifx\next\span %
            \global\advance\countREGISTER by 1%
            \ncase=0%
         \fi%
         \ifx\next| %
            \global\advance\countREGISTER by 2%
            \ncase=0%
         \fi
         \ifx\next\|
            \global\advance\countREGISTER by 2%
            \ncase=0%
         \fi
         \ifx\next\multispan
            \ncase=1%
            \global\advance\multispancount by 1%
         \fi
         \ifx\next\header
            \ncase=2%
         \fi
         \ifx\next\cr       \global\firstrowfalse \fi
         \ifx\next\endrow   \global\firstrowfalse \fi
         \ifx\next\crthick  \global\firstrowfalse \fi
         \ifx\next\crnorule \global\firstrowfalse \fi
         \ifx\next\crnoruleneg \global\firstrowfalse \fi
         \ifx\next\crthickneg  \global\firstrowfalse \fi
         \ifx\next\crneg       \global\firstrowfalse \fi
      \fi
\relax
      \ifcase\ncase %
         \let\next\COLcount%
      \or %
         \let\next\spancount%
      \or %
         \let\next\argCOLskip%
      \else %
      \fi %
   \fi%
   \next%
}%
\gdef\argROWskip#1{%
   \let\next\ROWcount \next%
}
\gdef\arghdROWskip#1{%
   \let\next\ROWcount \next%
}
\gdef\argCOLskip#1{%
   \let\next\COLcount \next%
}
}
}
\def\spancount#1{
   \nspan=#1\multiply\nspan by 2\advance\nspan by -1%
   \global\advance \countREGISTER by \nspan
   \let\next\COLcount \next}%
\def\dvr#1{\relax}%
\def\header#1{%
\dvr{1}{\let\cr=\@mpersand%
\hdtks={#1}%
\counthdROWS\hdtks\into\hdrows%
\advance\hdrows by 1%
\ifnum\hdrows=0 \hdrows=1 \fi%
\dvr{5}\makehdPREAMBLE{\the\hdrows}%
\dvr{6}\getHDdimen{#1}%
{\parindent=0pt\hsize=\hdsize{\let\ifmath0%
\xdef\next{\valign{\headerpreamble #1\crnorm}}}\dvr{7}\next\dvr{8}%
}%
}\dvr{2}}
\def\makehdPREAMBLE#1{
\dvr{3}%
\hdrows=#1
{
\let\headerARGS=0%
\let\cr=\crnorm%
\edef\xtp{\vfil\hfil\hbox{\headerARGS}\hfil\vfil}%
\advance\hdrows by -1
\loop
\ifnum\hdrows>0%
\advance\hdrows by -1%
\edef\xtp{\xtp&\vfil\hfil\hbox{\headerARGS}\hfil\vfil}%
\repeat%
\xdef\headerpreamble{\xtp\crcr}%
}
\dvr{4}}
\def\getHDdimen#1{%
\hdsize=0pt%
\getsize#1\cr\end\cr%
}
\def\getsize#1\cr{%
\endsizefalse\savetks={#1}%
\expandafter\lookend\the\savetks\cr%
\relax \ifendsize \let\next\relax \else%
\setbox\hdbox=\hbox{#1}\newhdsize=1.0\wd\hdbox%
\ifdim\newhdsize>\hdsize \hdsize=\newhdsize \fi%
\let\next\getsize \fi%
\next%
}%
\def\lookend{\afterassignment\sublookend\let\looknext= }%
\def\sublookend{\relax%
\ifx\looknext\cr %
\let\looknext\relax \else %
   \relax
   \ifx\looknext\end \global\endsizetrue \fi%
   \let\looknext=\lookend%
    \fi \looknext%
}%
%
%
\def\tablelet#1{%
   \tableLETtokens=\expandafter{\the\tableLETtokens #1}%
}%
\catcode`\@=12
%

%
%
%
%
%
%
%
%
%
%
%
%
\catcode`@=11

\font\seventeencp=cmcsc10 scaled\magstep3
\def\SIZE{\hsize=6.6truein\vsize=9.1truein}
\def\OFFSET{\voffset=1.2truein\hoffset=.8truein}
\def\papersize{\SIZE\OFFSET\skip\footins=\bigskipamount
\normaldisplayskip= 30pt plus 5pt minus 10pt}
\Pubnum={\rm CERN$-$TH.\the\pubnum }
\def\title#1{\vskip\frontpageskip\vskip .50truein
     \titlestyle{\seventeencp #1} \vskip\headskip\vskip\frontpageskip
     \vskip .2truein}
\def\author#1{\vskip .27truein\titlestyle{#1}\nobreak}

\def\p@bblock{\begingroup \tabskip=\hsize minus \hsize
   \baselineskip=1.5\ht\strutbox \topspace-2\baselineskip
   \halign to\hsize{\strut ##\hfil\tabskip=0pt\crcr
   \the \Pubnum\cr}\endgroup}
\def\makefootline{\iffrontpage\vskip .27truein\line{\the\footline}
                 \vskip -.1truein\line{\the\date\hfil}
              \else\line{\the\footline}\fi}
\paperfootline={\iffrontpage \the\Pubnum\hfil\else\hfil\fi}
\paperheadline={\iffrontpage\hfil
                \else\twelverm\hss $-$\ \folio\ $-$\hss\fi}
\newif\ifmref  
\newif\iffref  
\def\xrefsend{\xrefmark{\count255=\referencecount
\advance\count255 by-\lastrefsbegincount
\ifcase\count255 \number\referencecount
\or \number\lastrefsbegincount,\number\referencecount
\else \number\lastrefsbegincount-\number\referencecount \fi}}
\def\xrefsdub{\xrefmark{\count255=\referencecount
\advance\count255 by-\lastrefsbegincount
\ifcase\count255 \number\referencecount
\or \number\lastrefsbegincount,\number\referencecount
\else \number\lastrefsbegincount,\number\referencecount \fi}}
\def\xREFNUM#1{\space@ver{}\refch@ck\firstreflinetrue%
\global\advance\referencecount by 1
\xdef#1{\xrefend}}
\def\xrefend{\xrefmark{\number\referencecount}}
\def\xrefmark#1{[{#1}]}
\def\xRef#1{\xREFNUM#1\immediate\write\referencewrite%
{\noexpand\refitem{#1}}\begingroup\obeyendofline\rw@start}%
\def\xREFS#1{\xREFNUM#1\global\lastrefsbegincount=\referencecount%
\immediate\write\referencewrite{\noexpand\refitem{#1}}%
\begingroup\obeyendofline\rw@start}
\def\rrr#1#2{\relax\ifmref{\iffref\xREFS#1{#2}%
\else\xRef#1{#2}\fi}\else\xRef#1{#2}\xrefend\fi}
\referencecount=0
%
\space@ver{}\refch@ck\firstreflinetrue%
\immediate\write\referencewrite{}%
\begingroup\obeyendofline\rw@start{}%
\def\plb#1({Phys.\ Lett.\ $\underline  {#1B}$\ (}
\def\nup#1({Nucl.\ Phys.\ $\underline {B#1}$\ (}
\def\plt#1({Phys.\ Lett.\ $\underline  {B#1}$\ (}
\def\cmp#1({Comm.\ Math.\ Phys.\ $\underline  {#1}$\ (}
\def\prp#1({Phys.\ Rep.\ $\underline  {#1}$\ (}
\def\prl#1({Phys.\ Rev.\ Lett.\ $\underline  {#1}$\ (}
\def\prv#1({Phys.\ Rev. $\underline  {D#1}$\ (}
\def\und#1({            $\underline  {#1}$\ (}
\message{ by W.L.}
\everyjob{\input offset }
\catcode`@=12

\let\it=\sl

\def\KW{\rrr\KW{L. Krauss and F. Wilczek, Phys.Rev.Lett. 62 (1989) 1221.}}

\def\DIS{\rrr\DIS{T. Banks, \nup323 (1989) 90; \nextline
L. Krauss, Gen.Rel.Grav. 22 (1990) 50; \nextline
M. Alford, J. March-Russell and F. Wilczek, \nup337 (1990) 695;\nextline
J. Preskill and L. Krauss, \nup341 (1990) 50;\nextline
M. Alford, S. Coleman and J. March-Russell, \nup351 (1991) 735.
}}

\def\WH{\rrr\WH{For a review and references see: T. Banks, Physicalia
Magazine 12 (1990) 19.}}

\def\PR{\rrr\PR{J. Preskill, Ann.Phys. 210 (1991) 323.    }}

\def\BD{\rrr\BD{T. Banks and M. Dine, Phys.Rev.D45 (1992) 1424.}}

\def\KMN{\rrr\KMN{D. Kapetanakis, P. Mayr and H.P. Nilles,
\plb282 (1992) 95.}}

\def\M{\rrr\M{S.P. Martin, Gainesville preprint UFIFT-HEP   (1992).}}

\def\W{\rrr\W{P.L. White, Southampton preprint SHEP-91/92-24 (1992).}}

\def\SOO{\rrr\SOO{S.-J. Rey, Yale preprint YCTP-P14-92 (1992).}}

\def\CL{\rrr\CL{E. Chun and A. Lukas, Munich preprint TUM-TH-150/92
(1992).}}

\def\IRP{\rrr\IRP{L.E. Ib\'a\~nez and G.G. Ross, \nup368 (1992) 3.}}

\def\IR{\rrr\IR{L.E. Ib\'a\~nez and G.G. Ross, \plb260 (1991) 291.}}

\def\DF{\rrr\DF{E. D'Hooker and E. Farhi, \nup248 (1984) 59.}}

\def\PTWW{\rrr\PTWW{J. Preskill, J. Trivedi, F. Wilczek and
M. Wise, \nup363 (1991) 207.}}

\def\PVN{\rrr\PVN{P. van Nieuwenhuizen, Phys.Rep. 68 (1981) 191.}}

\def\BDA{\rrr\BDA{T. Banks and A. Dabholkar, Rutgers preprint
RU-99-09 (1992).}}

\def\KS{\rrr\KS{G. Lazarides and Q. Shafi, preprint BA-92-32 (1992);
\nextline L. Krauss and S.J. Rey, preprint YCTP-P9-92 (1992).}}

\def\GS{\rrr\GS{M. Green and J. Schwarz, \plb149 (1984) 117.}}

\def\LNS{\rrr\LNS{E. Witten, \plb149 (1984) 351;\nextline
W. Lerche, B. Nilsson and A.N. Schellekens, \nup299 (1988) 91;\nextline
M. Dine, N. Seiberg and E. Witten, \nup289 (1987) 585;\nextline
J. Atick, L. Dixon and A. Sen, \nup292 (1987) 109.}}

\def\DHVW{\rrr\DHVW{L. Dixon, J. Harvey, C. Vafa and E. Witten,
\nup274 (1986) 285.}}

\def\INQ{\rrr\INQ{L.E. Ib\'a\~nez, H.P. Nilles and F. Quevedo,
\plb187 (1987) 25.}}

\def\FINQ{\rrr\FINQ{A. Font, L.E. Ib\'a\~nez, H.P. Nilles and
F. Quevedo, \nup307 (1988) 109.}}

\def\CKN{\rrr\CKN{E.J. Chun, J.E. Kim and H.P. Nilles,
\nup370 (1992) 105.}}

\def\GW{\rrr\GW{S.L. Glashow and S. Weinberg, Phys.Rev.D15 (1977) 1958.}}

\def\I{\rrr\I{L.E. Ib\'a\~nez, CERN-TH.6501./92 (1992).}}

\def\E{\rrr\E{T. Eguchi, P. Gilkey and A. Hanson, Phys.Rep. 66 (1980)
214.}}

\def\CP{\rrr\CP{K.W. Choi, D. Kaplan and A. Nelson,
preprint UCSD-PTH-92-11 (1992);\nextline
M. Dine, R. Leigh and D. MacIntire, preprint SCIPP-92-16 (1992).}}

\def\OFFSET{\hoffset=12.pt\voffset=55.pt}
\def\SIZE{\hsize=420.pt\vsize=620.pt}

\catcode`@=12
\newtoks\Pubnumtwo
\newtoks\Pubnumthree
\catcode`@=11
\def\p@bblock{\begingroup\tabskip=\hsize minus\hsize
   \baselineskip=0.5\ht\strutbox\topspace-2\baselineskip
   \halign to \hsize{\strut ##\hfil\tabskip=0pt\crcr
   \the\Pubnum\cr  \the\Pubnumtwo\cr 
   \the\pubtype\cr}\endgroup}
\pubnum={6662/92}
\date{September  1992   }
\pubtype={}
\titlepage
\vskip -.6truein
\title{ More About Discrete Gauge Anomalies}
 \vskip 0.1truein
 \centerline{\bf Luis E. Ib\'a\~nez  *}
 \vskip .1truein
 \centerline{CERN, 1211 Geneva 23, Switzerland}
 \vskip .1truein
\abstract\noindent\nobreak

I discuss and extend several results concerning    the cancellation of
discrete gauge anomalies. I show how       heavy fermions
do not decouple in the presence of discrete gauge anomalies.
As a consequence, in general, cancellation of discrete gauge
anomalies cannot be described merely in terms of low energy operators
involving only the light
fermions. I also discuss cancellation of discrete gauge anomalies through
a discrete version of the Green-Schwarz (GS) mechanism as well as
the possibility of discrete gauge R-symmetries and their anomalies.
                   Finally, some      phenomenological applications
are discussed. This includes symmetries guaranteeing absence of FCNC
in   two-Higgs models and generalized matter parities stabilizing
the proton in the supersymmetric standard model. In the presence
of a discrete GS mechanism or/and gauge R-symmetries, new possibilities
for anomaly free such symmetries are found.

\bigskip
\bigskip
{\bf  *}\ Adress after September 1992: Departamento de Fisica
Te\'orica  C-XI, Universidad Aut\'onoma de Madrid, Cantoblanco,
28049, Madrid, Spain.

\endpage

\pagenumber=1
\sequentialequations

\leftline{\bf 1.\ Introduction }

\bigskip

Discrete symmetries are often impossed in Lagrangian field
theories for a variety of phenomenological motivations. Such kind
of symmetries seem to be required e.g. in different extensions
of the standard model like multi-Higgs or supersymmetric models
in order to avoid large flavour-changing neutral currents or
too fast proton decay. There is nothing wrong with discrete
symmetries except for the apparent lack of fundamental
(non-phenomenological) motivation for their existence and the
arbitrary number of existing possibilities. A further problem
for discrete symmetries may appear if indeed, as argued by some
authors,                  gravitational (e.g. wormhole \WH ) corrections
badly violate all non-gauge symmetries.

Those problems could be solved if the physically relevant discrete
symmetries were `gauge'  discrete       symmetries \KW ,\DIS ,\PR  .
Such discrete
gauge symmetries have been argued to be stable against large
gravitational corrections \KW ,\DIS             . Furthermore it was
pointed out in ref.\IR  that
                      discrete gauge symmetries are restricted by certain
anomaly cancellation conditions. Many candidate gauge discrete
symmetries may be ruled out on the basis of these conditions.
This has been applied in particular to the discrete symmetries
ensuring proton stability in the supersymmetric standard model
(SSM) \IRP , solar neutrino models  \KMN , \CL  and other
phenomenological questions         \M ,\W .
The possibility of interpreting CP as a gauge discrete symmetry
has also been considered in ref.\CP .

In the present paper I discuss further several aspects of the
cancellation of discrete gauge anomalies and clarify several issues
raised in refs.\IR ,\IRP\ and \BD .
In section two I emphasize a peculiar property of the discrete
gauge anomaly cancellation conditions which is that, in general,
heavy fermions
do not decouple in the presence of discrete gauge anomalies.
This is reminiscent of the work of ref.\DF  in which it was shown
that the top-quark does not decouple if one tries to make it
infinitely heavy in the standard model. This is particularly the
case of the cubic $Z_N$ discrete  gauge anomaly and the mixed
$Z_N$-gravitational anomaly which are sensitive to massive
fermions. In the case of the mixed $Z_N-SU(M)$ anomalies, the
contribution of light and heavy fermions to the anomaly cancels
separately  and hence the discrete anomaly is not sensitive to heavy
fermions. That is why                             the
$Z_N-SU(M)$ anomaly may be understood just in terms of the
relevant 't Hooft fermionic operator involving $only$ the
massless fields, a point of view which has specially been
emphasized in ref.\BD  .
                 On the other hand, the cubic $Z_N$ and the
mixed $Z_N$-gravitational anomalies are not easy to understand
from that point of view. Nevertheless, even though these two
anomalies depend on the heavy sector of the theory, they do it
in a definite manner which make possible to obtain usefull
constraints from them.

In section three I briefly discuss the cancellation of
discrete anomalies for the case of `gauge' R-symmetries.
This also provides us with an specific example of the problems
one may get if one tries to interpret $all$ discrete anomalies
in terms of a 't Hooft fermionic operator. The presence of discrete
gauge anomalies in four-dimensional strings is discussed in section
four. I also present several examples and describe     the way in  which
a discrete version of the Green-Schwarz mechanism operates \IRP ,\BD
(see also \SOO ).
In section five I consider a couple of phenomenological applications
including symmetries supressing flavour changing neutral currents
in two-Higgs models and symmetries supressing fast proton decay
in the supersymmetric standard model. In particular, I discuss
how the analysis in ref.\IRP   is modified   if
one considers i) discrete  R-symmetries   and/or ii) a discrete
Green-Schwarz mechanism. I find that in such a case new
`Generalized Matter Parities' could be anomaly free. However,
the presence either of gauged R-symmetries and/or a discrete
(GS) mechanism are only expected in theories with higher dimensions
like strings. Some final comments and comparison with other
approaches \PTWW\  are left for the sixth section.

\bigskip
\leftline{\bf 2.\ The Discrete Gauge Symmetry Anomaly Cancelation
Conditions}
\centerline{\bf and the Non-Decoupling of Heavy Fermions}

Let us start by briefly recalling the discrete gauge anomaly
conditions derived in ref.\IR .
                                                  Let us assume there is
an effective gauge $Z_N$ symmetry under which the massless fermions
of the theory transform with charges $q_j$.
There are several
types of ``discrete anomalies'' which must be cancelled by
appropriately choosing the chiral fermion content of the theory.
These are the following:

i) Cubic $Z_N^3$ anomaly cancelation condition:
$$\sum _i (q_i)^3\ = \ r\ N\ +\ \eta \  \ s\ {{N^3}\over 8}
\ \ ,\ \   s\in {\bf Z}. \eqn  \mastera $$
where $\eta =1,0$ for $N$=even, odd.

ii) Mixed $Z_N$-gravitational anomalies:
$$\sum _i(q_i)\ =\ r'\ N\ +\ \eta \ n'\ {N\over 2}\ \ ,\ n',r'\in{\bf Z}.
\eqn \masterb $$

iii) Mixed $Z_N$-$SU(M)$-$SU(M)$ anomalies:
$$\sum _iT_i\ (q_i)\ ={1\over 2}\ r''\ N \ \ ,\ r''\in {\bf Z}.
\eqn \masterc $$
Here $T_j$ is the quadratic $SU(M)$ Casimir corresponding to each
given representation (the normalization is such that the Casimir of
an $M$-plet is $=1/2$). The sums are over the light sector of the
theory (those fields which do not get a mass when the symmetry
breaking $U(1)\rightarrow Z_N$ takes place).

If          there are extra unbroken abelian factors, $U(1)_u$s,
in the theory one can
write down further constraints associated with mixed
$U(1)_u^2Z_N$ and $U(1)_uZ_N^2$ anomalies. However \IR\ they turn out
not to be very usefull without the precise knowledge of the
underlying theory before the symmetry breaking $U(1)\rightarrow Z_N$
and, particularly, the normalization of the $U(1)$'s. If that
normalization is known (e.g., if it is known that those $U(1)$'s
were embedded in some simple group), these anomalies are
restrictive of the massless fermion content and should be taken into
account.

Let us discuss some interesting aspects of the above
anomaly conditions.

{\bf i)}
   In the cubic  $Z_N^3$ anomaly conditions  \mastera\
above we have not indicated whether $r$ should be taken an integer
or not. In fact, it will always be an integer if the massive
fermions which gained masses in the process $U(1)\rightarrow Z_N$
have integer $Z_N$ charges.   It is easy to see \IR\  that the
contribution of massive fermions $\Psi _1^j,\Psi _2^j$ with
charges $Q_1^j,Q_2^j$ to the coefficient $r$ in \mastera\
is given by
$$
r_{\Psi }\ =\ -\ \sum _{j=heavies}\ p_j
(3{Q_1^j}^2\ -\ 3p_jNQ_1^j\ +\ p_j^2N^2)
\eqn \rpsi
$$
where the $p_j$ are integers obeying $Q_1^j+Q_2^j=p_jN$. These integers
$p_j$ carry information about the charges of the scalar fields which
break the symmetry $U(1)\rightarrow Z_N$. In
\mastera\ one has $r=r_{\Psi }+integer$.
It is then obvious that, if the massive fermions all have
integer charges, $r$ will be an integer \footnote*{Notice
that, for $N=3$, $r$ has to be in fact a multiple of 3.}.
In the absence
of
specific information about the massive sector, one cannot be sure
about the nature of $r$ which may perfectly be a fractional number.
In this case the cubic anomaly is not very usefull in
constraining low-energy physics \BD\ ,\IRP\  . On the other hand one can
reverse things and constraint the massive sector from
anomaly cancellation \BD ,\IRP . Indeed, since by definition
$\sum _i q_i^3$ is an integer, $N\times r_{\Psi }$ has to be an
integer which yields the consistency condition
$$
3N\ \sum _{j=heavies}\ p_jQ_1^jQ_2^j\ =\ integer
\eqn \conheavies
$$
where the sum  runs over all the massive pairs of fermions and
$Q_1^j+Q_2^j=p_jN$.
This equation should be valid for any normalization choice of
the massive fields and is a non-trivial constraint on the
possible charge assignements of massive fermions. Notice that
if the integer in eq. \conheavies\ is a multiple of $N$, the
massive fermions do not contribute to the cancellation of
the discrete anomaly and $r$ in eq.\mastera\ is an integer.

Eq. \conheavies\  has implications on the possible charges of
massive fermions. To be specific, consider a $Z_{N\times M}$
discrete gauge symmetry such that the massless fermions have
charges multiple of $M$. From the low energy point of view
only a $Z_N$ subgroup will be explicit in the interacting
Lagrangian, the massless fermions would have integer $Z_N$
charges and the massive fermions will in general
have fractional $Z_N$ charges proportional to $integer/M$.
The condition \conheavies\ will require
$$
{{3N}\over {M^2}}\  \sum _j\ a_j \ =\ a \ \ ,\ \ a=integer
\eqn \conN
$$
where the $a_j$ are integer coefficients. It is obvious from this
expression that $aM^2/3N$ has to be an integer. There are two cases
now depending whether $a$ is a multiple of $N$ or not. If $a$ is
a multiple of $N$, then one can immediately see that $r$ in \mastera\
is an integer also and       the anomaly cancellation condition
is equal to the one corresponding to integrally charged massive fermions.
If $a$ is not a multiple of $N$, then necessarily $M^2$ will have
to be a multiple of $N$, in order to $aM^2/3N$ being integer. Thus we
are left with two options concerning the cubic $Z_N^3$ anomaly,
either

{\bf 1)} The cubic anomaly condition \mastera\ is verified
with $r=integer$ or

{\bf 2)} There are massive fields with fractional $Z_N$ charges
proportional to $m/M$, $m\in {\bf Z}$,     with
$aM^2$ a multiple of $3N$.

This second case may be, in some cases, quite restrictive. Thus
consider a theory in which the massless fermions respect a
$Z_3$ symmetry ($N=3$). The minimum value of $M$ for which
condition \conN\ may be verified is $M=3$, meaning that the
complete theory has a discrete gauge symmetry as large as
$Z_9$.

The above shows us how the cubic $Z_N^3$ anomaly cancellation
condition \mastera\ with integer $r$ may be quite informative.
Either it is obeyed, or if not, it tells out something about
the charges of the massive fermions. On the other hand, there may
be cases in which we have extra information about the massive sector
(e.g. like some GUTs or string models) and in which
                    the integrality of the charges of the massive
fermions may be guaranteed from the  beginning.

{\bf ii)}
The mixed $Z_N$-$SU(M)$ anomaly conditions \masterc\  may also be
obtained from $SU(M)$ instanton considerations \BD     . This may be
easily understood from the 't Hooft fermionic operator
$$
 O_t\ =\ \prod _{lights} \psi \psi ....\psi
\eqn \tooft
$$
where the product extends over all the light fermions with
$SU(M)$ quantum numbers. The invariance of that operator
 with respect to the discrete gauge transformation  yields
the same result as eq. \masterc\ .

This type of
low energy Lagrangian interpretation does not seem very immediate
in the case of the cubic \mastera\ and gravitational \masterb\
anomaly conditions. The reason for that is clear in the analysis
from ref.\IR\  . The main reason for this difference is that the
contribution of the massive fermions with $SU(M)$ quantum numbers
to the mixed $Z_N-SU(M)$ anomaly vanishes identically, i.e., the
anomaly cancels separately both both massless and massive
fermions. In fact one should in principle multiply the operator
$O_t$ by a similar operator $O_T$ involving the heavy fermions
with $SU(M)$ quantum numbers. Since the discrete anomalies
for light and heavy fermions cancel separately this is
not necessary in this case.

This is in fact what is different in the other two type
of anomalies. In the case of the cubic anomaly the massive
fermions may contribute in two manners to the cancellation
of the light fields discrete anomalies. 1) the pairs of
massive fermions with Dirac masses may contribute if they
have fractional $Z_N$ charges by providing a fractional
$r$ coefficient in eq.\mastera\ ; 2) for even $N$ the
massive Majorana fields contribute the last term in
eq.\mastera  . In the case of the mixed $Z_N$-gravitational
anomaly, the contributions of the massive fermions with Dirac
masses cancel identically and only the massive Majorana fermions
may contribute to the cancellation of the light fields discrete
gauge symmetry (the last piece in eq.\masterb\ ).

It is obvious from the above discussion of   the cancellation
of the cubic $Z_N^3$ and mixed-gravitational discrete gauge
anomalies that {\it there is not decoupling of massive
fermions in                 a chiral theory with a massless
fermion sector  which has discrete gauge anomalies}.
It is also clear that one cannot understand $all$ the discrete
anomaly cancellation conditions merely in terms of instanton
physics involving just the massless fermions. Well on the contrary,
the mixed $Z_N$-$SU(M)$ anomalies are exceptional in that
the massive fermions do not contribute to anomaly cancellation,
and that is why such an interpretation is possible.
Concerning the mixed gravitational anomaly, in addition to
the fact that heavy Majorana particles do not in general
decouple from the low-energy physics, there are additional difficulties
for an instanton low energy interpretation because of the poor
knowlwdge of gravitational instanton physics. An example of these
difficulties will be shown in the next chapter while discussing
the mixed R-symmetry-gravitational anomalies.

The non-decoupling of massive fermions in the presence
of discrete gauge anomalies
     is reminiscent of the work in ref.\DF\ in which the
standard model in the limit of a heavy top quark is considered
           (more recent related work may be found in ref.\BDA ).
In that case the top quark never really decouples since then
one would be left with an anomalous theory. Instead, a Wess-Zumino
term is generated which helps to cancel the (perturbative)
anomalies. Since a top-less theory would also have a non-perturbative
$SU(2)$ anomaly one also adds a sort of `topological Wess-Zumino'
term which helps to cancel  that anomaly in the low energy sector.
In the case of the standard model, describing the Higgs field $\Phi $ in
terms of a unitary matrix $U$ such that $\Phi = \phi U$, this
non-perturbative Wess-Zumino term in the action has the form \DF
$$
\Delta S_1\ (U,A_{\mu},B_{\mu})\ =\cases{ 0  & U trivial in
$\Pi _4(SU(2))$ \cr   \pi & U non-trivial in $\Pi _4(SU(2))$ \cr}  \ .
\eqn \dels
$$

It is reasonable to conjecture that, since the discrete gauge
anomalies here considered are also non-perturbative, the effect
of the massive fermions in the low energy theory may also be
described by analogous `topological Wess-Zumino' terms in the
action. Consider first the case of the $Z_N^3$ cubic anomaly.
As we discussed, there are two contributions of the massive
fermions which help in cancelling the anomaly, one from the
massive Dirac pairs and the other from the Majorana massive fields.
The contribution of the latter is proportional to $nN^3/8, n\in {\bf Z}$
and is only present for even $N$. In fact one can see that this
contribution is only non-trivial for $N=2$. Indeed, one can write
$N=2N', N'\in {\bf Z}$. If $N'$ is odd one really has a discrete
$Z_2\times Z_{N'}$ symmetry and it is only the $Z_2$ piece which
has this term. If $N'$ is even, $nN^3/8=nN'^3$ is always a multiple
of $N=2N'$ and hence vanishes modulo $N$. Thus one only has to
worry about this term in a $Z_2$ cubic discrete anomaly. Concerning
the contribution from the massive Dirac pairs, it will appear
through a generation of a non-integer $r_{\psi }$ in eq.
\mastera\ (but such that $r_{\psi}N$ is integer), and may be
generically present for all $N$. Thus one can achieve
cancellation of discrete cubic anomalies from the light
fermions by formally adding in to the action terms  of the form
$$
\Delta S_{cubic}\ =\ \cases{ r_{\psi}2\pi   & N' even \cr
                r_{\psi }2\pi\ +\ n\pi & N' odd \cr}
\eqn \deldgs
$$
In the case of the mixed $Z_N$-gravitational anomaly, only massive
Majorana particles can contribute to the cancellation of the
discrete anomaly of the light fermions. One would expect then
a term of the form
$\Delta S_{grav} \ =\ n'\pi  $
which would be present only for even $N$ (but for any $N'\in {\bf Z}$).

\bigskip

\leftline{\bf 3. \ Discrete R-symmetry Anomalies }

In $N=1$ supersymmetric theories there are two type Abelian internal
symmetries: ordinary symmetries and R-symmetries. The first commute
with the SUSY generator and the second do not. It is then obvious
that one cannot gauge a continuous $U(1)$ R-symmetry since
supersymmetry would be broken. On the other hand there do in
general exist in certain cases `discrete  gauge R-symmetries'.
As remarked in ref.\IRP ,\BD\ , if the four-dimensional field theory
is obtained from a higher dimensional one (as may be the case in
Kaluza-Klein and string theories), there could be R-symmetries
which are discrete remnants of the rotation group in the compactified
dimensions. Indeed, in explicit 4-dimensional strings like
orbifold compactifications those discrete gauge R-symmetries exist.
They are `gauge' in the sense that they originate from the gauge
theory of          relativity in the extra dimensions.

In addition to that possibility, there is another type of
 continuous `gauged R-symmetries' which correspond to $U(1)$
symmetries in supersymmetric sigma models. In this case the gauge
connection does not propagate, it is instead an auxiliary vector
field which couples in a chiral manner to fermions. This sigma
model structure  with gauge connections coupling to R-symmetries
are in general present in $N=1$ supergravity Lagrangians. They
also appear in the particular class of $N=1$ supergravity
Lagrangians which correspond to the low energy limit of some
four-dimensional strings. In this particular case, these
gauged $U(1)$ R-symmetries can be broken down to a $Z_N$ R-symmetry
subgroup by the vev of some scalar field.

Using arguments similar to those in ref.\IR\  one can obtain the
equivalent discrete anomaly cancelation conditions for
discrete R-symmetries. A generic R-symmetry associates charges
for the gravitino and gaugino fields $Q^R_{3/2}=Q^R_{1/2}=  1$
and charges for the matter fermions $\psi _i$ equal to $q_i$.
In the process of symmetry breaking $U(1)^R\rightarrow Z_N$
(while preserving supersymmetry) some
matter fermions may get three types of masses: 1) Dirac mass terms
combining pairs of matter fields; 2) Majorana mass terms
3)   Dirac mass terms combining a matter field with a gaugino
field  as in the usual supersymmetric version of the Higgs effect.
Since, by assumption, supersymmetry is unbroken, there cannot be neither
gaugino Majorana mass terms nor mixed gravitino-gaugino nor
gravitino-matter fermion terms, since all of these would break
supersymmetry either explicitely (in the first case) or spontaneously
(in the last two cases).

Let us first consider the mixed discrete R-symmetry-$SU(M)$
anomaly. One can in fact directly apply     equation \masterc\ to
this case to obtain
$$
\sum _i T_i(q_i)\ +\ M\ =\ {1\over 2} r''\ N\ ,\ r''\in {\bf Z}
\eqn \rsun
$$
where $T_j$ is the quadratic $SU(M)$ Casimir corresponding to each
given representation and, again, the normalization is such that
the Casimir of an M-plet is $=1/2$. This expression may be
understood \BD\ by requiring that the corresponding instanton generated
operator in eq.\tooft\ vanishes.

The case of the mixed R-symmetry-gravitational anomaly is more
complicated since the gravitino field contributes to it.
The contribution of spin 3/2 and 1/2 fermions to the usual
gravitational chiral anomaly is given by \PVN\
$$
\int   d^4x  \sqrt{g}\ D_{\mu}J_5^{\mu}\ =\
{1\over {24}}\ \{ {1\over {16\pi ^2}}\ \int d^4x\sqrt(g)
R^{*}_{\mu \nu \rho \sigma }R^{\mu \nu \rho \sigma } \}
\  (21\ N_{3/2}\ -\ N_{1/2})
\eqn \rr
$$
where $N_{3/2},N_{1/2}$ are the number of spin 3/2 and spin 1/2
fermions. This tells us that the contribution of the gravitino
is (-21) times what a Weyl spinor contributes. This may be understood
counting     the number of fermion zero modes in the background of
a $K_3$ gravitational instanton \E\ . Apart from this, one can again use
eq.\masterb\ and obtain for the mixed $Z_N$-R-symmetry mixed
gravitational anomaly
$$
\sum _i\ q_i\ -\ 21\ +\ dim\ G\ =\ r'\ N\ +\ \eta \ n' \ {N\over 2}
\ \ , n',r'\in {\bf Z} \ \
\eqn \rgrav
$$
where $dim\ G$ is the dimension of the complete gauge group.
Notice                    that there is no obvious way by which
the above anomaly cancellation equation could be understood in
terms of an effective 't Hooft operator  involving just the
massless fermions. To start with, such a possible interpretation
would not be able to account for the last piece in eq.\rgrav\ ,
which represents the contribution of possible massive Majorana
fields to the cancellation of the anomaly. Secondly, it is not
obvious what fermion operator might account for the factor 21
between gravitino and matter fields.
Notice that our eq.\rgrav\  does not agree with the result
presented  in ref.\BD\  which apparently did not
include the correct zero mode counting.

\bigskip

         {\bf 4.\ Discrete Gauge Symmetries in String Models and
the Discrete Green-Schwarz Mechanism}

In four-dimensional string theories the presence of discrete
symmetries is ubiquitous. Many of them are Abelian $Z_N$ symmetries
which have a gauge character. It is perhaps usefull to enumerate a
few examples of discrete gauge symmetries appearing in string
models.

i) In four-dimensional strings constructed a la Gepner, the models
are obtained by tensoring $N=2$ superconformal minimal models
both for right and left movers in such a way that the internal
central charge is $c=9$. Each of the  superconformal blocks
presents a $Z_{k+2}$ symmetry, where $k$ is the level of the
minimal model. If there are n such factors the resulting model
has a $\Pi _i{Z_{k_i+2}}_R\otimes \Pi _i {Z_{k_i+2}}_L $, $i=1,..n$
discrete symmetry, although often only a subgroup of this
symmetry is realized in the massless particles. Some of these
are normal symmetries whereas others are discrete $R$-symmetries.
These models also have $n-1$  $U(1)$ gauge bosons, the so-called
enhanced $U(1)$s. One can easily check in these models the presence
of some massless singlet scalars (moduli), which, if given a
non-vanishing vev, break the $U(1)$s down to some of the
$Z_{k_i+2}$ symmetries described above. Thus plenty of the discrete
symmetries in these models may be understood as discrete gauge
symmetries.

2) Discrete gauge symmetries are also abundant in orbifold
four-dimensional strings. For example, the discrete symmetries
associated to the $Z_N$ or $Z_N\times Z_M$ point group twists
of the six-dimensional tori may be understood as discrete gauge
symmetries. Indeed, orbifolds have enhanced $U(1)$ symmetries
at some multicritical points of the untwisted moduli (e.g., the
$E_6\times E_8\times SU(3)$ symmetry of the standard $Z_3$
orbifold is enhanced by extra $U(1)^6$ interactions). The untwisted
moduli have integer charges under these enhanced $U(1)$s whereas
the twisted fields have generically fractional charges (i.e.,
multiple of $1/N$ for a $Z_N$ twist). If the untwisted  moduli
get non-vanishing vevs, the enhanced $U(1)$s are spontaneously
broken to $Z_N$ subgroups which are then discrete gauge symmetries.

3)  The two above examples correspond to `enhanced' $U(1)$
symmetries breaking down to some abelian discrete gauge group. There
are also plenty of examples of gauge groups belonging to the
`$E_8\times E_8$' sector which break also to subgroups containing
a gauge $Z_N$ subgroup. Let us give here a $Z_3$ orbifold example
which will also serve us to               briefly describe the
Green-Schwarz \GS\ mechanism often at work in four-dimensional strings.
Take one of the four modular-invariant $(0,2)$ embeddings of
the $Z_3$ orbifold, the one leading to the gauge group
$E_7\times SO(14)\times U(1)_1\times U(1)_A$. This model may be
obtained \DHVW ,\INQ\
         by embedding the $Z_3$ twist in the $E_8\times E_8$
degrees of freedom through a shift  ${\vec V}=1/3(110...0)\times
(20...0)$. The massless charged chiral fields of the model
are the following: three copies of
$(56,1;0,1)+(1,1;0,-2)+(1,14;-2,0)+(1,64;1,0)$ in the untwisted
sector; 27 copies of $(1,14;-2/3,2/3)+(1,1;4/3,-4/3)$ in the
twisted sector and 81 copies of $(1,1;4/3,2/3)$ from twisted
oscilators. One interesting feature of this model is that
the ABJ gauge anomalies of the $U(1)_A$ Maxwell field do not
cancell, as the reader may easily check. In fact, this is not quite
true since the anomaly is cancelled by the four-dimensional version \LNS\
of the Green-Schwarz (GS) mechanism. In a simplified way this mechanism
works as follows. In four-dimensional strings the gauge
kinetic piece of the Lagrangian has the generic form
$$
\phi (x)\sum _ik_iF^2_i\ +\ i\eta (x)\sum _ik_iF_i{\tilde F}_i\ ,
\eqn \gs
$$
where $\eta (x)$ is the axionic field which is the partner of the
string dilaton $\phi (x)$ and the sum runs over the different
gauge groups of the model. The $k_i$ are the Kac-Moody levels of
the different gauge groups. They are integers for non-Abelian
groups but may be non-integer numbers for the  $U(1)$s.
The GS mechanism works by assigning a non-trivial transformation
behaviour of the field $\eta (x)$ under a gauge transformation
$V^{\mu}_A\rightarrow V^{\mu }_A+\partial ^{\mu }\theta (x)$
of the anomalous $U(1)_A$:
$$
\eta (x)\ \rightarrow \ \eta (x)\ -\ \theta (x)\ \delta _{GS}
\eqn \gsm
$$
where $\delta _{GS}$ is a constant. If the anomaly coefficients
$C_i$ corresponding to the triangle anomalies involving one $U(1)_A$
gauge boson and any other pair of gauge bosons are in the
ratio
$$
 {{C_1}\over {k_1}}\ =\ {{C_2}\over {k_2}}\ =.....\ = \ \delta _{GS}\ ,
\eqn \ccc  $$
the corresponding anomalies will be cancelled by the $U(1)_A$
gauge variation of the second term in eq.\gs .
It is easy to check that in the above $Z_3$ orbifold example the
anomaly coefficients are in the apropriate ratio and the
anomalies are cancelled by this mechanism. One can also see \FINQ\
that there is a flat direction in the scalar potential in
which one of the $(56,1;0,1)$ massless fields get a vev and
break the symmetry:
$$
E_7\times U(1)_A\ \rightarrow \ E_6\ \times {Z_3}^A \ ,
$$
i.e. the $U(1)_A$ symmetry is broken to a $Z_3$ gauged subgroup.
Under this ${Z_3}^A$ symmetry all untwisted fields are singlets
whereas all twisted states transform with the phase $exp(i4\pi/3)$,
as the reader may easily check by looking to the $U(1)_A$ charges
of the fields in the model.

                                     All the above string
examples verify the discrete gauge anomaly cancellation
conditions. In fact, in the case of the $Z_3$ orbifold examples,
since the multiplicity of the charged fields is always a multiple
of 3, the anomalies cancel  in a trivial way. The same is
true for the ${Z_3}^A$ example above, even though the original
$U(1)_A$ is  `anomalous', the residual ${Z_3}^A$ is not, due
to the multiplicity of the twisted fields. In general, if an
`anomalous' $U(1)_A$  is broken to a subgroup ${Z_N}^A$, the latter
will be `anomalous' only if
$${{C_i}\over {k_i}}\ =\ \delta _{GS}\ \not= \ 0\ ,\ mod\ N \ .
\eqn \ano
$$
A model with this property is provided by the $Z_3$ orbifold
example of  ref.\CKN\  . This is a model with three discrete Wilson
lines in which the gauge group is
$SU(5)\times SU(3)\times SU(2)\times U(1)^9$ and one of the $U(1)$s
is `anomalous'. The mixed anomalies of this $U(1)$ with the
non-Abelian gauge interactions are respectively
$C_5=C_3=C_2=-44$ which is $\not= 3$ mod 3.  The model has a flat
direction in the untwisted sector (involving scalar fields corresponding
to the same complex plane and with $SU(3)\times SU(2)\times U(1)_A$
quantum numbers $(3,2;-3), ({\bar 3},1;0)$ and $(1,2;+3)$).
When these fields get equal vevs the symmetry breaking $U(1)_A\rightarrow
Z_3$ occurs. This residual $Z_3$ is thus apparently `anomalous' but
in fact this anomaly is cancelled by an apropriate discrete shift
of the axion field $\eta (x)$. This `discrete Green-Schwarz
mechanism', as its continuous counterpart, is likely to be
present in many specific four-dimensional strings. Another example
in a  Gepner-type construction with Wilson lines  was given in
ref. \BD  .

\bigskip

\leftline{\bf 5.\ Phenomenological Applications}

Discrete symmetries are often used in order to constrain Lagrangians
for a variety of phenomenological problems. An important question is
whether there is any reason why low energy discrete symmetries
should be gauged. I can think of three main motivations for that
assumption. 1) Gauge discrete symmetries are stable against
gravitational (e.g., wormholes) corrections which have been
argued          violate all global symmetries. 2) Most (may be all)
discrete symmetries of the effective Lagrangian in 4-dimensional
strings may be understood as gauge discrete symmetries.
3) The usual theoretical prejudice in favour of local versus
global symmetries. Assuming that the low energy discrete symmetries
come somehow from a larger gauge symmetry gives a rationale for the
very existence of discrete symmetries. On the contrary, the existence
of discrete global symmetries seem unmotivated from a fundamental
point of view.

If one accepts that low energy discrete symmetries are gauge, they
must obey the anomaly cancellation conditions discussed in chapter
2. These equations then constrain the possible allowed discrete
symmetries in a substantial manner. It makes then sense to check
what are the discrete symmetries used in the most popular
phenomenological Lagrangians and check for discrete gauge
anomaly cancellation conditions. Examples of relevant discrete
symmetries are the following:

{\bf 1)}   Discrete symmetries supressing flavour changing
neutral currents in two-Higgs models.

                  These are needed in generic extensions of the
standard model involving more than one Higgs doublet. Consider
the two Higgs-doublet case and    a           generation-blind $Z_N$
symmetry acting on the five independent standard model fermion
multiplets $Q,u,d,L,e$ and the two Higgs doublets with opposite
hypercharges $H,\bar H$  as
$$
\Psi _j\ \rightarrow \ e^{i\alpha _j2\pi/N}\ \ \Psi _j\ \
j=Q,u,d,L,e,H,\bar H \ \ .
\eqn \stand
$$
where the $\alpha _j$ are
integers.
The absence of flavour-changing neutral currents (FCNC) is
guaranteed if the Higgs which couples to the charge 2/3 quarks
is different from the one which couples to the charge 1/3 quarks
and the leptons \GW\ . Consider the three independent
$Z_N$ generators  discussed in ref.\IRP
$$
R_N\ =\ e^{i2\pi I_3^R/N}\ \ ,\ \
A_N\ =\ e^{i2\pi Y_A/N}\ \ ,\ \
 L^j_N\ =\ e^{i2\pi L^j/N}\ \     \eqn \gen  $$
where $j=1,2,3$ correspond to the three lepton numbers
\footnote*{ We will later on denote the equivalent overall
lepton number generator by $L$.}
The form
of the charge       generators $I_3^R, Y_A, L^j$ is shown in
table 1. The invariance of the standard model Lagrangian with
respect to $I_3^R$ and $L^j$ correspond to the invariance of
the $renormalizable$ Lagrangian with respect to Baryon and Lepton
numbers. If one adds a second Higgs doublet, one has to imposse some
symmetry in order to ensure the absence of FCNC, as we remarked
above. It is easy to check that the most general discrete
symmetry which works may be written as
$$
S_{  N }\ =\ R_N^m\times (L_N^j)^{p_j} \times A_N     \ \  ,
\eqn \sfcnc
$$
i.e., essentially the discrete $A_N$ generator, up to discrete
baryon or lepton number discrete generators.

Following the above argumentation, one should imposse that the
 $S_N$ symmetry should be anomaly-free. The mixed $S_N$-$SU(3)$,
and $SU(2)$            anomaly coefficients are given by \I
$$
\eqalign{C_3\ &=\ - {{N_g}\over 2} \cr
C_2\ &=\ - {{N_g}\over 2}\ -\ {{\sum _j^{N_g} p_j}\over 2} \cr }
\eqn \ccc
$$
where $N_g$ is the number of quark-lepton generations.
The coefficients $C_3,C_2$ should vanish modulo $N/2$ for those
anomalies to cancel, and this only happens for $N=Ng=3$.
Thus {\it the unique anomaly free $S_N$ symmetry guaranteeing
absence of flavour changing neutral currents is $A_3$}. The
other possibilities are, either anomalous or related to
$A_3$ by discrete baryon or lepton number rotations.
Notice that the mixed $A_3$-gravitational anomaly yields
$C_{grav}=N_g(-3-2)=-15=0$ mod 3 and hence it also vanishes.

If one allows for the existence of a discrete Green-Schwarz
mechanism along the lines of the previous sections, new possibilities
appear. In fact, as shown in ref.\I\ , in this case a {\it continuous
$U(1)_A$ symmetry coupling to the $Y_A$ generator} in table  1
may be gauged and its mixed anomalies with the standard model
interactions may be cancelled through a continuous Green-Schwarz
mechanism. Correspondingly, discrete symmetries $A_N$, for any
$N$, may be gauged if the corresponding mixed anomalies are
cancelled by a discrete Green-Schwarz mechanism. For example,
looking at eqs.\ccc\ one sees that the condition $C_3=C_2, mod\ N/2$
for discrete GS mechanism to work yields the condition
$\sum _j ^{N_g} p_j=0\ mod\ N$ and no constraint on $  m$.
(Here we have assumed the standard values $k_3=k_2$. Using different
k's would allow for even more possibilities).
We thus conclude that, in the absence of a discrete GS mechanism,
$S_3$ symmetries are the unique anomaly-free symmetries supressing
FCNC. In the presence of a discrete GS mechanism, all discrete
$S_N$ symmetries in eq.\sfcnc\  satisfying $\sum ^jp_j=0$ mod $N$
may in principle be gauged. Notice also that the number of
possibilities may be further restricted e.g., if one also
imposses cancellation of mixed gravitational anomalies.
Of course, this type of analysis may be easily extended to
models with a more complicated Higgs structure.

In ref.\PTWW\ an anomalous global $Z_2$ symmetry corresponding to
a   $S_2$ generator was considered. It is clear from the above
discussion that this symmetry may be made anomaly free (and hence
be gauged) if one includes a discrete GS mechanism as above.

{\bf 2)}  Generalized Matter Parities in the Supersymmetric
Standard Model

A general analysis of the possible generation-independent
discrete gauge symmetries ensuring proton stability in the
supersymmetric standard model (SSM) was presented in ref.\IRP .
The result of the analysis shows that, assuming the particle
content of the SSM, there are only a few discrete $Z_N$ {\it generalized
matter parities} which are anomaly free.
In terms of the generators in
eq.\gen\ one can write the   flavour-independent discrete symmetries as
\IRP
$$
g_N\ =\ R_N^m\ \times \ A_N^n\ \times L_N^p
\eqn \gmp
$$
The finite list of such anomaly free symmetries is as
follows:

i) A $Z_3$ called `baryon parity' $B_3$ which is generated by
$B_3=R_3L_3$. The phase assignements of matter fields with
respect to this symmetry are $g_{(Q,u,d,L,e)}=
(1,\alpha ^2,\alpha ,\alpha ^2,\alpha ^2)$, where
$\alpha =exp(i2\pi /3)$. This is the only symmetry which is
anomaly-free just with the massless fermion content, without
the need to include a possible effect from massive fermions.
It also forbids the presence of dangerous dimension five operators
which may give rise to fast proton decay.

ii) The standard $Z_2$ R-parity. This is equivalent to the
symmetry generated by $R_2$. In this case the mixed gravitational
anomaly    only cancels if there are massive Majorana fermions
(behaving like right-handed neutrinos) which help to cancel
the anomaly (last piece in eq.\mastera\ ).

iii) There are three other $Z_3$ symmetries which can be made
anomaly-free if there are massive fermions with fractional
$Z_3$ charges. They are the symmetries generated by
$R_3$, $L_3$ and $R_3L_3^2$. These symmetries seem to obey
all the anomaly-cancellation conditions \mastera\ -\masterc\ .
However, as indicated in the footnote in section 2,  for
$Z_3$ symmetries $r$         must be a multiple of 3
(and $rN$ a multiple of 9) if the massive particles have integer
$Z_3$ charges. This is the case for $B_3$, but not for these
other three symmetries. This implies that, for these three symmetries
to be anomaly-free, there must be massive fractionally charged
fermions helping in cancelling the anomaly. As discussed in
section 2,  this means that the actual symmetry of the theory
is at least as large as $Z_9$ and not just merely $Z_3$.

This is the finite list of anomaly free  possibilities for the particle
content of the $minimal$ SSM.
The   three symmetries in iii)   were overlooked in the preprint version
of ref.\IRP and included in the published version. Although less
atractive than $B_3$ and $R_2$ because they involve larger order
symmetries, they cannot be discarded merely on anomaly
cancellation grounds.

If we enlarge the spectrum by considering an axion-dilaton
system making possible a discrete version of the GS mechanism,
the number of potentially anomaly-free solutions
substantially increases. Let us consider just the mixed
$Z_N$-$SU(3)$ and $SU(2)$ anomalies for the moment.
For a discrete GS mechanism to work one needs
to have (see eq.\ccc )
$$
\eqalign{ -n\ N_g\ &=\ 2\ \delta _{GS}\ k_3   \cr
        -n\ N_g\ -p\ N_g\ +\ n\ N_D\ &=\ 2\ \delta _{GS}\ k_2
\ \ \ ,\ mod\ N \cr }
\eqn \gisa
$$
where $N_g$ is the number of generations and $N_D$ the
number of Higgs pairs. $\delta _{GS} $ is the Green-Schwarz term
coefficient which is a model dependent constant and $k_2,k_3$ are
the levels of the groups. Normally one assumes $k_2=k_3$ (it is
required in order     to obtain reasonable coupling constant unification
predictions) and then, substituting $\delta _{GS}$ one gets
after some trivial algebra
$$
p\ -\ n{{N_D}\over {N_g}}\ +\ t\ {N\over {N_g}}\ =\ 0\ \ ,\ \
t\in {\bf Z}
\eqn \pnor
$$
and again no constraint on $m$ (the $I_3^R$ generator has
no mixed $SU(2)$,$SU(3)$ anomalies). It is easy to find solutions
to these equations: The $Z_2$ symmetries generated by $A_2L_2$
($m=0,n=p=1$) and $R_2A_2L_2$ ($m=n=p=1$) can be made anomaly-free
through this mechanism. For $Z_3$ symmetries one has solutions
only for $n=0$ mod 3 and hence no $new$ symmetries can be made
anomaly-free through the GS mechanism. On the other hand,
and unlike the case without GS mechanism, $Z_N$ symmetries
with $N>3$ may be made anomaly free. Examples are the
$Z_4$ symmetry generated by $A_4^{-1}L_4$ and the
$Z_5$ symmetry generated by $A_5  L_5^2$.

In addition to eq.\gisa\ one can further imposse cancellation of
mixed gravitational anomalies. The equivalent condition is now
$$
-(5n\ +\ p\ -\ m)\ N_g\ +\ 2nN_D\ +\ {1\over 2}\eta sN\ =\ 24
\  \delta _{GS} \ \ mod \ N\ , s\in {\bf Z}
\eqn \gisagrav
$$
where $\eta =1,0$ for $N=$ even, odd.
The factor 24 accounts for the graviton normalization
and plays the same role as the $k_i$ do for gauge fields.
Identifying the $\delta _{GS}$ coefficient with those in
eq.\gisa yields further constraints. In particular, the
two symmetries $A_2L_2$ and $R_2A_2L_2$ are anomaly free whereas the
above $Z_5$ example is not.

We thus observe  that the existence of a dilaton-axion
multiplet with transformation properties as
required by the discrete GS mechanism increases
                the number of possibilities. However,
since the natural setting for the existence of such a
mechanism is string theory, one might consider any
evidence in favour of any of these new symmetries as
indirect evidence for string theory.

One point not discussed in ref.\IRP\ was the anomaly
cancellation constraints on the {\it generalized R-matter
parities}, i.e., discrete R-symmetries protecting the
SSM from too fast proton decay. Since string and other
higher dimensional theories R-symmetries may also be
gauge in origin, it is worth checking how effective
are the anomaly cancellation conditions for these type
of R-symmetries. The most general discrete R-symmetry of
the SSM can be writen as
$$
{\tilde g}_N\ =\ {\tilde P}_N\times R_N^m\times A_N^n\times \prod _j
{L_N^j}^{p_j}
\eqn \tila
$$
where $j$ runs over lepton numbers and ${\tilde P}_N$ is defined by
${\tilde P}_N(d\theta )$ $=exp(i2\pi /N)$ $(d\theta )$, $\theta $ being
the superspace Grassman variable
            \footnote*{Notice that the definition of ${\tilde P}_N$ here
is different from that of $P_N$ in ref.[6]  by a factor of 2 in the
exponent.
This is done so that the action on the fermions is generically a
$Z_N$ and not a $Z_{2N}$ symmetry. Note e.g. that the $P_2$ symmetry
in ref.[6]  is in fact a $Z_4$ symmetry whereas ${\tilde P}_2$ here
is a genuine $Z_2$.}.
              On the other hand, the action on the Higgs fields $H,
{\bar H}$ is defined in a slightly different way so that the usual
Yukawa couplings are always allowed (see table 1)
Let us       consider to start with the mixed R-symmetry-$SU(3)$
and $SU(2)$ anomalies from a generic generation-independent
R-symmetry
$$
\eqalign{ 6\ -\ 4\ N_g\ -\ n\ N_g\ &=\ 0\ \ \ mod\ N   \cr
4\ -\ 4\ N_g\ -\ (n+p)\ N_g\ +\ N_D\ (2+n)\ &=\ 0\ mod\ N  \cr }\ .
\eqn \rssm
$$
For the case of the minimal SSM one has $N_g=3,N_D=1$ and those two
equations reduce to $3p-n=0$ and $6+3n=0$ mod N, with arbitrary $m$.

There is no non-trivial $Z_2$ generalized matter R-parity which is
anomaly free (${\tilde P}_2$ is trivial since it corresponds to
a $Z_2$ fermion number).
                     On the other hand, one-third of the
$Z_3$ R-symmetries  (eight of them) are anomaly-free. We list them
here for convenience of the reader:${\tilde P}_3,{\tilde P}_3R_3$,
${\tilde P}_3L_3,{\tilde P}_3L_3^2$,
${\tilde P}_3L_3R_3,{\tilde P}_3L_3R_3^2$,
${\tilde P}_3L_3^2R_3,{\tilde P}_3L_3^2R_3^2$.
One can further impose the cancellation of mixed gravitational
anomalies eq.\rgrav\  which yields
$$
-21\ +\ 12\ +\ N_g(-15-5n-p+m)\ +\ N_D(4+2n)\ =\ 0\ \ mod\ 3  \ .
\eqn \congrav
$$
This equation has no solution for the realistic case with
$N_g=3, N_D=1$. Thus, unless one adds extra singlets to the
minimal SSM, there are no anomaly free $Z_3$ R-symmetries
stabilizing the proton. On the other hand, extra singlets
are likely to exist anyhow in supergravity theories so that
one should not rule out completely the above symmetries.
Unlike
the case of standard discrete  generalized matter parities, there
are also plenty of solutions for $N$ higher than 3.
Thus, e.g., the $Z_4$ symmetry ${\tilde P}_4L_4^2A_4^2$ has no mixed
$SU(2)$ and $SU(3)$ anomalies. On the other hand the ${\tilde P}_4$
symmetry (equivalent to the $P_2$ of ref.\IRP ) is anomalous.
The number of possibilities may be reduced by impossing also the
cancellation of mixed R-symmetry-gravitational anomalies as in
the $Z_3$ cases above.  This I leave as an exercise to the reader.
Additional anomaly-free R-symmetries may be found if
one allows for cancellation of anomalies through
a discrete GS mechanism involving the R-symmetries.

As a general conclusion one sees that the possibilities
for anomaly-free symmetries and R-symmetries stabilizing the
proton are               increased if 1) one allows for
a discrete GS mechanism and 2) one allows for gauged  R-symmetries.
For $N<4$, however, only a couple of new $Z_2$ symmetries,
which correspond to $A_2L_2$ (a `lepton parity') and $R_2A_2L_2$ (a
`baryon parity') may be made anomaly free
(a discrete GS mechanism is required). A discrete GS mechanism
would also be required to get new anomaly free R-symmetries for
$N<4$.

   The above new anomaly free symmetries    are only
really expected in theories with higher dimensions
like strings. In some way, any evidence for this type of symmetries
would be an indirect evidence for higher dimensions.
                      In theories without higher dimensions
only the few $Z_2,Z_3$ symmetries described at the  begining of this
section are anomaly free. Out of those only the baryon-parity
$B_3$ is anomaly-free without the help of contributions from
hypothetical heavy fermions.

\bigskip

\leftline{\bf 6.\ Final Comments}

In the approach described here the cancellation of discrete
gauge symmetries is used as a constraint on low energy physics.
A different approach has been considered in ref.\PTWW  and \KS .
These authors consider $anomalous$ (but, in general, non-gauge)
discrete symmetries. If these discrete symmetries are anomalous,
the QCD anomaly may give rise to interesting effects which could be
of use in solving the domain-wall problem present in some models.
This is an interesting scenario which could be present e.g. in
many grand unified theories. However, as I argued above, it is
reasonable to assume that the discrete symmetries are gauge
in order to protect the symmetries from large gravitational effects.
Furthermore, a gauge origin for discrete symmetries gives a rationale
for their very existence. String theories, as I described above,
have plenty of discrete gauge symmetries and all the examples found
up to now are anomaly free. This is why I believe impossing the
constraint of anomaly cancellation is an interesting constraint.
This seems an approach orthogonal to that of ref.\PTWW and \KS \ which
make use of anomalous discrete symmetries, a situation not expected
in string models.

In the case in which a discrete GS mechanism is at work, if one
ignores the axion-dilaton sector, the rest of the massless fermionic
sector looks anomalous, although the overall theory is anomaly-free.
It is in this sense \BD\  that one can say that anomalous discrete
symmetries may appear in string models. However, it is not clear that
one can use the arguments of \PTWW\ for this kind of `anomalous'
discrete symmetries since the effects of the dilaton-axion sector
cannot in general be neglected at high temperatures. Furthermore,
the discrete symmetry will also $necessarily$ have mixed $SU(2)_L$
anomalies whose non-perturbative effects at finite temperature also
have to be included.
                                             All these points
would have to be taken into account in trying to use the
philosophy of ref.\PTWW\ to these `pseudoanomalous' discrete symmetries
appearing in some string models with a discrete GS mechanism.

\endpage

\par \penalty-400 \vskip\chapterskip
   \spacecheck\referenceminspace \immediate\closeout\referencewrite
   \referenceopenfalse
   \line{\fourteenrm\hfil REFERENCES\hfil}\vskip\headskip
   \input referenc.texauxil

\endpage

\bigskip
\bigskip

\begintable
   \
 \|         $Q$ | $u$     |$d$      |     $L$ |   $e$
\| ${\tilde H}$        |${\tilde {\bar H }}$
|   ${\tilde g},{\tilde W},{\tilde B}$                              \cr
     $R$    \| 0 | -1| 1 | 0 | 1 \| -1| 1 | 0              \cr
$Y_A$\|  0      |  0       |  -1     | -1      | 0       \| 1| 0  | 0 \cr
  $L_i$    \| 0        | 0        | 0   | -1 | 1 \|  0  |  0 |  0  \cr
${\tilde P} $ \| -1  |   -1  |  -1  |  -1  | -1    \| 1  | 1  |  1
\endtable

\bigskip
\bigskip
\centerline{Table\ 1.\ Generators of        symmetries in the SM and SSM}

\vfill\eject\bye